\documentclass[12pt]{article}

\usepackage{amsmath,amssymb,graphicx} 
\usepackage{epsf}
\usepackage{pstricks}
\usepackage{cite}

\newcommand{\beq}{\begin{eqnarray}}
\newcommand{\eeq}{\end{eqnarray}}

\newcommand{\centeron}[2]{{\setbox0=\hbox{#1}\setbox1=\hbox{#2}\ifdim
                                        
\wd1>\wd0\kern.5\wd1\kern-.5\wd0\fi
\copy0

\kern-.5\wd0\kern-.5\wd1\copy1\ifdim\wd0>\wd1
                                       \kern.5\wd0\kern-.5\wd1\fi}}
\newcommand{\ltap}{\>\centeron{\raise.35ex\hbox{$<$}}
                               {\lower.65ex\hbox{$\sim$}}\>}
\newcommand{\gtap}{\>\centeron{\raise.35ex\hbox{$>$}}
                               {\lower.65ex\hbox{$\sim$}}\>}

\newcommand\ZZ{\hbox{\zfont Z\kern-.4emZ}}
\font\zfont = cmss10 
\newcommand{\sfrac}[2]{{\textstyle\frac{#1}{#2}}}

\def\tv#1{\vrule height #1pt depth 5pt width 0pt}

\begin{document}
\begin{titlepage}
\begin{flushright}
{\tt hep-ph/0505001}
\end{flushright}

\vskip.5cm
\begin{center}
{\huge \bf Top and Bottom: \\
\vspace{.2cm}
a Brane of Their Own 
}

\vskip.1cm
\end{center}
\vskip0.2cm

\begin{center}
{\bf
{Giacomo Cacciapaglia}$^{a}$, {Csaba Cs\'aki}$^{a}$,
{Christophe Grojean}$^{b}$,\\
{Matthew Reece}$^{a}$, {\rm and}
{John Terning}$^{c}$}
\end{center}
\vskip 8pt

\begin{center}
$^{a}$ {\it Institute for High Energy Phenomenology\\
Newman Laboratory of Elementary Particle Physics\\
Cornell University, Ithaca, NY 14853, USA } \\
\vspace*{0.1cm}
$^{b}$ {\it Service de Physique Th\'eorique,
CEA Saclay, F91191 Gif--sur--Yvette, France} \\
\vspace*{0.1cm}
$^{c}$ {\it
Department of Physics, University of California, Davis, CA  
95616.} \\
\vspace*{0.3cm}
{\tt  cacciapa@mail.lns.cornell.edu, csaki@lepp.cornell.edu,
grojean@spht.saclay.cea.fr, mreece@mail.lns.cornell.edu,  terning@physics.ucdavis.edu}
\end{center}

\vglue 0.3truecm

\begin{abstract}
\vskip 3pt
\noindent
We consider extra dimensional descriptions of models where there are two 
separate strongly interacting sectors contributing to electroweak symmetry breaking (``topcolor'' type 
models). In the extra dimensional picture there would be two separate (anti-de Sitter) bulks meeting on 
the Planck brane, with each bulk having its own corresponding IR (TeV) brane. Sources for electroweak symmetry 
breaking can then be localized on both of these IR branes, while the different generations of fermions
may be separated from each other. We describe the modes propagating in such a setup, and consider the cases
where the electroweak symmetry breaking on either of the two IR branes come either from a higgsless scenario 
(via boundary conditions) or a (top-)Higgs. We show that the tension that exists between obtaining a large top quark mass and the 
correct value of the $Zb\bar{b}$ couplings in ordinary higgsless models can be largely relieved in the 
higgsless---top-Higgs versions of the two IR brane models. This may also be true in the purely higgsless---higgsless 
case, however since that model is necessarily in the strongly coupled regime the tree-level results for the properties
of the third generation may get large corrections. A necessary consequence of such models is the appearance 
of additional pseudo-Goldstone bosons (``top-pions''), which would be strongly coupled to the third generation. 
\end{abstract}

\end{titlepage}

\newpage


\section{Introduction}
\label{sec:intro}
\setcounter{equation}{0}
\setcounter{footnote}{0}
There has been a tremendous explosion of new models of electroweak symmetry breaking
including large extra dimensions \cite{ADD}, Randall-Sundrum
\cite{RS}, gauge field Higgs
\cite{A5Higgs}, gauge extensions of the minimal supersymmetric standard model (MSSM) \cite{gaugextension}, little Higgs
\cite{littleHiggs}, and the fat Higgs
\cite {fatHiggs}.  All of these new models share one feature in common: a light Higgs.
With the realization that unitarity can be preserved in an extra-dimensional model by
Kaluza-Klein (KK) towers of gauge fields rather than a scalar \cite{otherunitarity,CGMPT}
a more radical idea has emerged: higgsless models~\cite{CGPT,Nomura,BPR,CGHST}. 
The most naive implementations
of these models face a number of phenomenological challenges, mostly related to avoiding strong coupling while satisfying bounds from precision
electroweak measurements \cite{BPR,DHLR1,BN,CCGT,DHLR2,BPRS,MSU,corrections,Howard,Maxim}.  
Collider signatures of higgsless models have been studied in~\cite{BMP,fermioncollider}, further
discussions on unitarity can be found in~\cite{Papucci,unitarityfermion,otherunitarity_2}, while other ideas
related to higgsless models can be found in~\cite{spurion,Nick,Otherhiggsless}.
However it has gradually emerged that, in a slice of 5 dimensional anti de Sitter space (AdS$_5$) with
a large enough curvature radius and light fermions almost evenly distributed in the bulk, 
$WW$ scattering is perturbative because the KK gauge bosons can be below 1 TeV, and the $S$ parameter and most other experimental constraints are satisfied because the coupling
of the light fermions  to the KK gauge bosons is small \cite{CuringIlls,MSUdeloc,otherdelocalizedfermions}.  
The outstanding problem
is how to obtain a large enough top quark mass without messing up the left-handed top and bottom
gauge couplings or the $W$ and $Z$ gauge boson masses themselves.  The tension arises~\cite{ADMS,ACP}
 because in order to get a large mass it would seem that the top quark must be close to the TeV brane
where electroweak symmetry is broken by boundary conditions. However this implies that
the top and (hence) the left-handed bottom have large couplings to the KK gauge bosons, and thus
have large corrections to their gauge couplings.  Furthermore this arrangement leads to a large
amount of isospin breaking in the KK modes of the top and bottom which then feeds into the $W$ and 
$Z$ masses through vacuum polarization at one loop~\cite{ADMS}.

It was previously suggested  \cite{CuringIlls} that a possible solution to this problem would be for the third generation
to live in a separate AdS$_5$.  In terms of the AdS/conformal field theory (CFT) correspondence~\cite{Maldacena,APR,RZ} this means that the
top and bottom (as well as $\tau$ and $\nu_\tau$) would couple to a different (approximate) 
CFT sector than the one which provides
masses to the $W$ and $Z$  as well as the light generations.  There is a long history of models
where the mechanism of electroweak symmetry breaking is different for the third generation.  This
is most often implemented as a Higgs boson that couples preferentially to the third generation (a.k.a. a ``top-Higgs")
but has also appeared in other guises such as top-color-assisted-technicolor (TC$^2$) where
top color \cite{Hill:topcolor} produces the top and bottom quark masses and technicolor produces all the other masses. From the point of view of AdS/CFT, double CFT sectors have been considered for a variety of reasons.
The setup is usually taken to two slices of AdS$_5$ ``back-to-back" with a shared Planck brane.
This is intended to approximately describe the situation of two strongly coupled CFT's that both couple to the same weakly coupled sector such as would arise when two conifold singularities are near each other in a higher dimensional space.
The tunneling between the two AdS ``wells" was considered in \cite{Dimopoulos:2001ui}
in order to generate hierarchies. More recently  inflationary models \cite{inflation} have been based on one or
more AdS wells. For other extra dimensional implementation of topcolor-type models 
see~\cite{ACD,otherxdimtopcolor}.

In this paper we will consider models of electroweak symmetry breaking with two  ``back-to-back" AdS$_5$'s in detail.
The motivation for these models is to be able to separate the dynamics responsible for the large top quark mass
from that giving rise to most of electroweak symmetry breaking. Thus we will assume that the light fermions
propagate in an AdS$_5$ sector that is essentially like the higgsless model described in~\cite{CuringIlls}, while the
third generation quarks would propagate in the new AdS$_5$ bulk. To analyze such theories we first discuss in detail
what the appropriate boundary and matching conditions are in such models. Then we consider the different 
possibilities for electroweak symmetry breaking on the two IR branes (Higgs---top-Higgs, higgsless---top-Higgs and 
higgsless---higgsless) and derive the respective formulae for the gauge boson masses. We then discuss the CFT interpretation 
of all of these results. Also from the CFT interpretation we find that there have to exist uneaten light
pseudo-Goldstone bosons (``top-pions'') in this setup. This is due to the fact that doubling
the CFT sectors implies a larger global symmetry group, while the number of broken gauge symmetries remains unchanged. 

After the general discussion of models with two IR branes we focus on those that can potentially solve the issues related
to the third generation quarks in higgsless models. A fairly simple way to eliminate these problems is by considering
the higgsless---top-Higgs case, that is when most of electroweak symmetry breaking originates from the higgsless sector,
but the top quark gets its mass from a top-Higgs on the other TeV brane (which also gives a small contribution to the 
$W$ mass ). The only potential issue is that since we assume the top-Higgs vacuum expectation value (VEV) $v_t$ on this brane to be significantly smaller than 
the Standard Model (SM) Higgs VEV, the top Yukawa coupling needs to be larger and thus non-perturbative. Also, the coupling of the 
top-pions to $t\bar{t}$ and $t\bar{b}$ will be of order $m_t/v_t$. Ideally, one would like to also eliminate the top-Higgs 
sector arising from the new TeV (IR) brane. In this case in order to get a very heavy top one needs to take the 
IR cutoff scale on the new side much bigger than on the old side ($\sim $ TeV), while keeping the top and bottom sufficiently far away from the 
new IR brane (in order to ensure that the bottom couplings are not much corrected). However, to make sure that 
most of the contributions to electroweak symmetry breaking are still coming from the old side one needs to choose a
smaller AdS radius for the side where the top lives. In this case perturbative unitarity in $WW$ scattering is 
still maintained. However, it will also imply that the new side of the 5D gauge theory is strongly coupled for all energies.
Electroweak precision observables are shielded from these contributions by at least a one-loop electroweak suppression,
however it is not clear that the KK spectrum of particles mostly localized on the new side would not get order one 
corrections and thus modify results for third generation physics significantly.

Finally we analyze some phenomenological aspects of this class of models.
An interesting prediction is the presence of a scalar isotriplet (top-pions), and eventually a top-Higgs.
The top-pions get a mass at loop level from gauge interactions, and the mass scale is set by the cutoff scale on the new TeV brane so that they can be quite heavy.
The main feature that allows us to distinguish such scalars from the SM or MSSM Higgses is that they couple strongly with the top (and bottom) quarks, but have sensibly small couplings with the massive gauge bosons and light quarks.
They are expected to be abundantly produced at the Large Hadron Collider (LHC), via the enhanced gluon or top fusion mechanisms.
If heavy, the main decay channel is in (multi) top pairs, although the golden channel for the discovery is in $\gamma \gamma$ and $Z Z$.
Thus, a heavy resonance in $\gamma \gamma$ and $l^+ l^- l^+ l^-$, together with an anomalously large rate of multi-top events, would be a striking hint for this models.

\section{Warmup: Boundary Conditions for a $U(1)$ on an Interval}
\label{sec:warmup}
\setcounter{equation}{0}
\setcounter{footnote}{0}

As an introduction to the later sections, we will in this section first present
a discussion on  how to obtain the boundary conditions (BCs) for a 
$U(1)$ gauge group on an interval, broken on both ends by two localized Higgses. The major focus will be on
 explaining  the effects 
of the localized Higgs fields on the boundary conditions for the $A_5$ bulk field, and possible effects of mixings
among the scalars, and the identification of possible uneaten physical scalar fields. 
We will also be allowing here a very general gauge kinetic function ${\cal K}(z)$ in the bulk, which will 
mimic both the effect of possible warping, and also the presence of a Planck brane separating two bulks with different
curvatures and gauge couplings. We use a $U(1)$ so that we keep the discussion
as simple as possible, while in the later sections we will use straightforward generalizations of the results obtained here 
for more complicated groups.

We assume that without the Higgs field being turned on the 
BCs are Neumann for the $A_\mu$ and Dirichlet for the $A_5$, as in the usual orbifold projection. This is a possible 
BC allowed by requiring the boundary variations of the action to be vanishing~\cite{CGMPT}.
With this choice, before the Higgs VEVs are turned on, there is no zero mode for the $A_5$ and all 
the massive degrees of freedom are eaten by the massive vector KK modes. In order to be able to clearly
separate the effects of the original boundary conditions from those of the localized fields added on the boundary,
we will add the localized fields a small distance $\epsilon$ away from the boundary. Of course later on we will be 
taking the limit $\epsilon \to 0$.

Thus the Lagrangian we consider is:
\begin{equation}
\mathcal{L} = \int_{L_1}^{L_2}\, dz\, \left\{ - \mathcal{K}(z) \frac{1}{4 g_5^2} F_{M N}^2+ \mathcal{L}_1 \delta (z-L_1-\epsilon) + \mathcal{L}_2 \delta (z-L_2+\epsilon) \right\}\,.
\end{equation}
As explained above, the generic function $\mathcal{K}$ encodes both the eventual warping of the space and a possible $z$-dependent 
kinetic term corresponding to a different $g_5$ on the two sides of the Planck brane.
We also assume that any eventual discontinuity is regularized such that $\mathcal{K}$ is continuous and non-vanishing.

The localized Lagrangians are the usual Lagrangians for the Higgs field in 4D~\footnote{Note that warp factors are usually added in the localized lagrangians, so that the scale $v$ is naturally of order $1/R$, and $\lambda$ of order 1. Such factors are not relevant for the discussion at this point, so we will neglect them for the moment.} ($i=1,2$):
\begin{equation}
\mathcal{L}_i = \left| \mathcal{D}_\mu \phi_i \right|^2 - \frac{\lambda_i}{2} \left( \left| \phi_i \right|^2 - \frac{1}{2} v_i^2 \right)^2
\end{equation}
and they will induce non vanishing VEVs $v_i$ for the Higgses, around which we expand:
\begin{equation}
\phi_i = \frac{1}{\sqrt{2}} \left( v_i + h_i \right) e^{i \pi_i/v_i}\,.
\end{equation}
The above Lagrangian contains some mixing terms involving $A_\mu$ that we want to cancel out with a generalized $R_\xi$ gauge fixing term. Expanding up to bilinear terms:
\begin{multline} \label{eq:lagrexp}
\mathcal{L} = \int_{L_1}^{L_2}\, dz\, \left\{ \frac{\mathcal{K}(z)}{g_5^2} \left( -\frac{1}{4} F_{\mu \nu}^2 + \frac{1}{2} \left( \partial_z A_\mu \right)^2 +  \frac{1}{2} \left( \partial_\mu A_5 \right)^2 - \partial_\mu A_5 \partial_5 A^\mu \right)\right. \\   
+\left[ \frac{1}{2} \left( \partial_\mu h_1 \right)^2 - \frac{1}{2} \lambda_1 v_1^2 h_1^2 +\frac{1}{2} \left( \partial_\mu \pi_1 -  v_1 A_\mu \right)^2 + ... \right] \delta (z-L_1-\epsilon) \\
\left.+\left[\frac{1}{2} \left( \partial_\mu h_2 \right)^2 - \frac{1}{2} \lambda_2 v_2^2 h_2^2 + \frac{1}{2} \left( \partial_\mu \pi_2 - v_2 A_\mu \right)^2 + ... \right] \delta (z-L_2+\epsilon) \right\}\,.
\end{multline}

The crucial point now is the integration by parts of the mixing term in the bulk.
As a consequence of the displacement of the localized Lagrangians $\mathcal{L}_{1,2}$, the integral splits into three regions limited by the regularized branes where the Higgs interactions are localized.
The contributions of the edges vanish in the limit $\epsilon \rightarrow 0$, so that the BC on the true boundaries are effectively 
``screened'', and a mixing between $A_\mu$ and $A_5$ on the branes is generated:
\begin{multline}
\left( \int_{L_1}^{L_1+\epsilon} + \int_{L_1+\epsilon}^{L_2-\epsilon} + \int_{L_2-\epsilon}^{L_2} \right) \, dz \, \mathcal{K} A_5 \partial_5 \partial_\mu A^\mu = \\
\left( \int_{L_1}^{L_1+\epsilon}  + \int_{L_2-\epsilon}^{L_2} \right) \, dz (...) - \int_{L_1+\epsilon}^{L_2-\epsilon}\,dz\, \partial_5 \left( \mathcal{K} A_5 \right) \partial_\mu A^\mu +\left[\mathcal{K} A_5 \partial_\mu A^\mu \right]_{L_1+\epsilon}^{L_2-\epsilon}\,.
\end{multline}
Note that the boundary terms would vanish on the true boundaries $L_{1,2}$, 
however they don't vanish on the branes at $L_1+\epsilon$ and $L_2-\epsilon$,
and thus the minimization of the action will require that $A_5$ has to be non-zero on the branes.
In other words, in the limit $\epsilon \rightarrow 0$ the $A_5$ field will develop a discontinuity on the boundaries.

We can now add a bulk and two brane gauge fixing Lagrangians\footnote{In~\cite{Muck} the authors considered a similar situation, but adding gauge fixing terms in the KK basis. As it has to be expected, our approach leads to the same results.}:
\begin{multline}
\mathcal{L}_{GF} = \int_{L_1}^{L_2} dz \left\{ - \frac{1}{g_5^2} \frac{1}{2 \xi} \left( \partial_\mu A^\mu - \xi \partial_z \left( \mathcal{K} A_5 \right) \right)^2 
- \frac{1}{2 \xi_{1}} \left( \partial_\mu A^\mu + \xi_{1} \left( v_{1} \pi_{1} - \frac{\mathcal{K}}{g_5^2} A_5 \right) \right)^2 \cdot \right.\\
 \delta (z-L_1) - \left. \frac{1}{2 \xi_{2}} \left( \partial_\mu A^\mu + \xi_{2} \left( v_{2} \pi_{2} + \frac{\mathcal{K}}{g_5^2} A_5 \right) \right)^2 \delta (z-L_2) \right\}\,,
\end{multline}
where the three gauge fixing parameters are completely arbitrary, and the unitary gauge is realized in the limit where all the $\xi$'s are sent to infinity.\footnote{Note that we have chosen a bulk gauge fixing term with a different $z$ dependence than the bulk gauge kinetic term, i.e., we have included a $z$ dependence in the gauge fixing parameter $\xi$. This allows us to obtain a simple equation of motion for $A_5$, at the price of a non-conventional form of the gauge propagator that will not be well suited for a warped space loop calculation in general $\xi$ gauge. In the unitary gauge that we will use in this paper, all the $\xi$ dependence vanishes (and thus also the $z$ dependence of $\xi$ will be irrelevant).} This gauge fixing term is devised such that 
all the mixing terms between $A_\mu$ and $A_5,\pi_{1,2}$
cancel out. The full Lagrangian then leads to the following equation of motion for $A_\mu$ (in the unitary gauge):
\begin{equation} \label{eom:Amu}
\frac{1}{\mathcal{K}} \partial_z \left( \mathcal{K} \partial_z A_\mu + m^2 A_\mu\right) =0\,,
\end{equation}
while the  BCs, fixed by requiring the vanishing of the boundary variation terms in
Eq.~(\ref{eq:lagrexp}), are:
\begin{equation}
\frac{\mathcal{K}(L_{1,2})}{g_5^2} \partial_z A_\mu \mp v_{1,2}^2 A_\mu =0\,.
\end{equation}

The bulk equation of motion for the scalar field $A_5$ will result in:
\begin{equation}
\partial_z^2 \left(\mathcal{K} A_5 \right) + \frac{m^2}{\xi} A_5  =  0 \ ,\label{eom:A5}
\end{equation}
and the $\pi$'s obey the following equations of motion on the branes:
\begin{equation} \begin{array}{c}
\left( \frac{m^2}{\xi_{1}} - v_{1}^2 \right) \pi_{1} + v_1 \frac{\mathcal{K}(L_1)}{g_5^2} \left. A_5 \right|_{L_1} =  0\,, 
\\
\left( \frac{m^2}{\xi_{2}} - v_{2}^2 \right) \pi_{2} - v_2 \frac{\mathcal{K} (L_2)}{g_5^2} \left. A_5 \right|_{L_2}  =  0 \,.
\end{array}\label{eom:pi}
\end{equation}
These last equations fix the values of $\pi_{1,2}$ in terms of the boundary values of $A_5$.
Finally, requiring that the boundary variations of the full action with respect to $A_5$ vanish, combined with the 
above expression (\ref{eom:pi}) for the $\pi$'s, will give the desired 
BCs for $A_5$:
\begin{equation} 
\left\{ \begin{array}{l}
\displaystyle \left. \partial_z \left( \mathcal{K} A_5 \right) - \frac{\xi_{1}}{\xi} \frac{\mathcal{K}(L_1)}{g_5^2} \frac{m^2/\xi_{1}}{m^2/\xi_{1} - v_{1}^2} A_5\right|_{L_1} = 0\,, \\
\displaystyle \left. \partial_z \left( \mathcal{K} A_5 \right) + \frac{\xi_{2}}{\xi} \frac{\mathcal{K} (L_2)}{g_5^2} \frac{m^2/\xi_{2}}{m^2/\xi_{2} - v_{2}^2} A_5 \right|_{L_2}= 0\,.
\end{array} \right. \end{equation}

 From Eq.~(\ref{eom:pi}) one can see that $\pi$ is not independent of $A_5$.
In the unitary gauge ($\xi \rightarrow \infty$) it is also clear from Eq.~(\ref{eom:A5}) that all the massive modes of $A_5$
are removed. This is simply expressing the fact that $A_5$ and the $\pi$'s are the sources of the longitudinal components
of the massive KK modes, and will be eaten. The only possible exception for the existence of a physical mode is if there is 
a massless state in $A_5$. Without the Higgses on the boundaries this would not be possible due to the Dirichlet BC.
However, we have seen above that the BC for the $A_5$ is significantly changed in the presence of the localized Higgs, and now
a massless state is possible. Physically, this expresses the fact that there are ``enough'' modes in $A_5$ to provide all the 
longitudinal components for the massive KK modes. If we add some localized Higgs fields, then there may be some 
massless modes left over uneaten. We will loosely refer to such modes as pions, to emphasize that these are physical 
(pseudo-)Goldstone bosons.
In the case of a massless physical pion mode, the BCs for $A_5$ simplify to:
\begin{equation}
\left. \partial_z \left( \mathcal{K} A_5 \right)\right|_{L_1\,\mbox{\tiny and}\, L_2} = 0\,.
\end{equation}

The solution to the equation of motion for the zero mode is of the form:
\begin{equation}
A_5 = \frac{g_5^2 d}{\mathcal{K} (z)},
\end{equation}
and using (\ref{eom:pi}) we also get that  
\begin{equation}
\pi_1 = \frac{d}{v_1}\,, \quad \pi_2 = - \frac{d}{v_2}\,.
\end{equation}
As expected in the higgsless limit (namely $v_i \rightarrow \infty$), the $\pi$'s vanish.
However, a massless scalar is still left in the spectrum if one chooses to break the gauge symmetry on both ends of the 
interval.
It is also interesting to note that in the limit where the function $\mathcal{K}$ is discontinuous, the solution $A_5$ develops a discontinuity as well.
But, the function $\mathcal{K} A_5$ is still continuous, so that no divergent term appears in the action for such a solution.

Finally, the spectrum also contains two scalars localized on the branes, corresponding to the physical Higgs fields.
As in the usual 4D Higgs mechanism, they will pick a mass proportional to the quartic coupling, 
$m^2_{h 1,2} = \lambda v_{1,2}^2$ (and decouple in the higgsless limits $v_i \to \infty$).

\subsection{Double AdS case}

\begin{figure}[tb]
\begin{center}
\includegraphics[width=11cm]{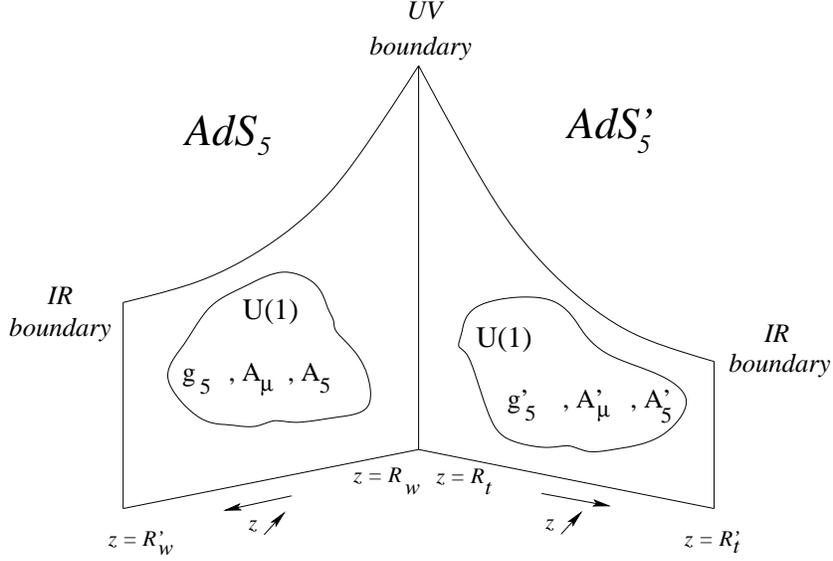}
\end{center}
\caption{Schematic view of the double AdS space that we consider.} \label{fig:setup}
\end{figure}

The physical setup we are actually interested in consists of two AdS$_5$ spaces intersecting along a codimension 
one surface (Planck brane) that would serve as a UV cutoff of the two AdS spaces. The whole picture can be seen as two Randall-Sundrum (RS) models glued together along their Planck boundary, as in Fig.~\ref{fig:setup}.
The two AdS spaces are characterized by their own curvature scale, $R_w$ and $R_t$. 
We define two conformal coordinate systems on the two spaces, namely ($i=w,t$):
\beq
ds^2=  \left( \frac{R_i}{z}\right)^2   \Big( \eta_{\mu \nu} dx^\mu dx^\nu - dz^2 \Big)\,.
\eeq
The common UV boundary is located at the point $z=R_i$ in the coordinate system associated to each brane. Each AdS space is also cut by an IR boundary located respectively at $z=R_w^\prime$ and $z=R_t^\prime$.

Alternatively, we can also think of the two AdS spaces as an interval with boundaries given by the IR branes, and the Planck brane as a singular point in the bulk.
  From this point of view we can apply the formalism developed above, in particular
we can define the function $\mathcal{K}$ in the two spaces:
\begin{equation}
\mathcal{K} = \left\{ \begin{array}{lcl}
\displaystyle \frac{R_w}{z\vphantom{{Z_Z}_Z}} & \mbox{for}& R_w \leq z \leq R_w^\prime\,,\\
\displaystyle \frac{g_5^2\vphantom{{Z^Z}^Z}}{{g_5'}^2} \frac{R_t}{z'}& \mbox{for}& R_t \leq z' \leq R_t^\prime\, ,
\end{array}\right. \end{equation}
where we have allowed for a different value of the bulk gauge coupling on the two sides if  $g_5 \neq g_5'$.
In order to maintain the traditional form of the metric in both sides of the bulk, we have chosen a peculiar coordinate 
system where $z$ is growing from the Planck brane towards both the left and the right. This way most formulae from RS 
physics will have simple generalizations, however it will also imply some unexpected extra minus signs.
Note that if $g_5 = g_5'$, the function $\mathcal{K}$ is continuous on the Planck brane. 
However, a discontinuity is generated if we define different gauge couplings on the two sides.
As noted before, the only effect of such choice will be a discontinuity in the wave function of the scalar field $A_5$ on the Planck brane.

The Lagrangians localized on the two IR branes that cut the two spaces are ($i=w,t$):
\begin{equation} \label{eq:Llocgen}
\mathcal{L}_i = \left(\frac{R_i}{R'_i}\right)^2 \left\{ \left| \mathcal{D}_\mu \phi_i \right|^2 -\left(\frac{R_i}{R'_i}\right)^2 \frac{\lambda_i}{2} \left( \left| \phi_i \right|^2 - \frac{1}{2} v_i^2 \right)^2 \right\}\,,
\end{equation}
where, introducing the above warp factors, all the scales and constants have natural values, namely $\lambda_i \sim 1$ and $v_i \sim 1/R_i$.

For the vectors, Eq.~(\ref{eom:Amu}) reduces to the usual RS equation of motion in the two bulks, with BCs given by the mass terms on the 
respective IR branes:
\begin{eqnarray}
\left.\partial_z A_\mu(z) \right|_{R_w^\prime} + \frac{g_5^2}{R_w} \frac{(v_w R_w)^2}{R_w^\prime} A_\mu(R_w^\prime) &=&0\,,\\
\left.\partial_z A'_\mu(z) \right|_{R_t^\prime} + \frac{{g'_5}^2}{R_t} \frac{(v_t R_t)^2}{R_t^\prime} A'_\mu(R_t^\prime) &=&0\,.
\end{eqnarray}
The equation of motion (\ref{eom:Amu}) is satisfied on the Planck brane as well, so we can translate it into matching conditions  for the solutions in the two bulks.
In particular, assuming that there are no interactions localized on the Planck brane, Eq.~(\ref{eom:Amu}) implies that the functions $A_\mu(z)$ and $\mathcal{K}(z) \partial_z A_\mu(z)$ are continuous:\begin{eqnarray}
\displaystyle A_\mu (R_w) & = & A'_\mu (R_t)\,, \\
\displaystyle \frac{1}{g_5^2}\left. \partial_z A_\mu (z)\right|_{R_w} & = & - \frac{1}{{g_5'}^2}\left. \partial_z A'_\mu (z)\right|_{R_t}\,. \label{eq:matchDAmu}
\end{eqnarray}  
The minus sign in the derivative matching Eq.~(\ref{eq:matchDAmu}) comes from the coordinate systems that we are using.
Note that the condition (\ref{eq:matchDAmu}) would be modified by localized terms: for instance, if we add a mass term for $A_\mu$, generated by a localized Higgs, the matching becomes:\beq
\displaystyle \frac{1}{g_5^2}\left. \partial_z A_\mu (z)\right|_{R_w} + \frac{1}{{g_5'}^2}\left. \partial_z A'_\mu (z)\right|_{R_t} + v_{Planck}^2 A_\mu = 0\,. \label{eq:matchDAmuMass}
\eeq
In particular, in the large VEV limit, it is equivalent to the vanishing of $A_\mu$ (and of $A_\mu'$).

Let us next consider the equations determining possible massless scalar pion modes.
As discussed above, from Eq.~(\ref{eom:A5}) we can read off that the continuity condition has to be applied to the functions $\mathcal{K}(z) A_5(z)$ and $\partial_z (\mathcal{K}(z) A_5(z))$:
\begin{eqnarray}
\displaystyle\frac{1}{g_5^2} A_5(R_w) & = &\frac{1}{{g_5'}^2} A_5'(R_t)\,, \\
\displaystyle \frac{R_w}{g_5^2}\left. \partial_z \frac{A_5(z)}{z}\right|_{R_w} & = & - \frac{R_t}{{g_5'}^2}\left. \partial_z \frac{A_5'(z)}{z}\right|_{R_t}\,.
\end{eqnarray}  
The solution then is:
\begin{equation} \begin{array}{rcl}
A_5 (z) & = & g_5^2 d \frac{z}{R_w}\,,\\
A'_5 (z') & = & {g_5'}^2 d \frac{z'}{R_t}\,,
\end{array} \quad \begin{array}{rcl}
\pi_w & = & - \left(\frac{R_w^\prime}{R_w}\right)^2 \frac{d}{v_w}\,, \\
\pi_t & = & - \left(\frac{R_t^\prime}{R_t}\right)^2 \frac{d}{v_t}\,.
\end{array} \end{equation}

Finally, the two Higgses localized on the IR branes get a mass proportional to the quartic couplings in the potentials:
\begin{equation}
m_{h_w}^2 =  \lambda_w \frac{(v_w R_w)^2}{R_w^{\prime 2}}\,, \quad  m_{h_t}^2 = \lambda_t \frac{(v_t R_t)^2}{R_t^{\prime 2}}\,.
\end{equation}
Their masses are naturally of order $R_w^{\prime -1}$ or $R_t^{\prime -1}$ respectively, and have an upper bound given by the breakdown of perturbative unitarity.
Such limits will be much looser than in the SM, due to the contribution of the gauge boson KK modes, as we will see in the following sections.

\section{The Standard Model in Two Bulks: Gauge Sector}
\label{sec:SM}
\setcounter{equation}{0}
\setcounter{footnote}{0}

We will now use our general formalism developed above to put
the Standard Model in two AdS$_5$ bulks. 
Our goal will be to separate electroweak
symmetry breaking from the physics of the third generation fermions. In each
AdS bulk, we have a full $SU(2)_L\times SU(2)_R\times U(1)_{B-L}$ gauge
symmetry, so that a custodial symmetry is protecting~\cite{ADMS} the $\rho$ parameter. 
We want to specify the boundary and matching conditions according to the 
following symmetry breaking pattern: on the common UV brane only  
$SU(2)_L\times U(1)_Y$ survives, while on the IR boundaries two Higgses break the gauge group to the $SU(2)_D\times U(1)_{B-L}$ subgroup.

The UV brane matching conditions arise from considering an
$SU(2)_R$ scalar doublet with a $B-L$ charge 1/2, that acquires a
VEV $(0,v)$. As discussed above, all of the
gauge fields $A^L_\mu$, $A^R_\mu$, and $B_\mu$ are continuous. 
The Higgs, in the limit of large VEV, forces the gauge bosons of the broken generators to vanish on the Planck brane:
 $A^{R1,2}_\mu = 0$ and $B_\mu - A^{R3}_\mu = 0$, thus breaking
$SU(2)_R \times U(1)_{B-L}$ to $U(1)_Y$. 
On the other hand, on the unbroken gauge fields $A^{La}_\mu$ and ($g_{5L}, g_{5R}, \tilde{g}_5$ are the 5D gauge couplings of SU(2)$_L$, SU(2)$_R$ and U(1)$_{B-L}$)
\beq 
B^Y_\mu = \frac{g_{5R}^2 \tilde{g}_5^2}{g_{5R}^2 + \tilde{g}_5^2}
\left(\frac{1}{g_{5R}^2} A^{R3}_\mu + \frac{1}{\tilde{g}_5^2} B_\mu \right)\,,
\eeq 
we need to impose the continuity condition in Eq.~(\ref{eq:matchDAmu}).
The complete set of Planck brane matching conditions then reads
\begin{eqnarray}
&{\rm at }\ 
\begin{array}{l}
z=R_w\\
z'=R_t
\end{array}
&
\left\{
\begin{array}{l}
A^{L\,a}_\mu = A^{\prime\, L\,a}_\mu\, , 
A^{R\,a}_\mu = A^{\prime\, R\,a}_\mu\, ,
B_\mu  = B^\prime_\mu \, ,
\\
\tv{15}
A^{R\, 1,2}_\mu =0 \, , 
B_\mu - A^{R\,3}_\mu = 0 \,, 
\\
\tv{15}
\partial_z  A^{L\,a}_\mu + 
\partial_z  A^{\prime\,L\,a}_\mu =0\, ,
\\
\tv{15}
\frac{1}{g^2_{5R}} \partial_z A^{R\,3}_\mu  + \frac{1}{g^2_{5R}}  \partial_z A^{\prime\,R\,3}_\mu 
+ \frac{1}{{\tilde g}^2_5} \partial_z B_\mu + \frac{1}{{\tilde g}^2_5} \partial_z B^\prime_\mu
=0\, .
\end{array}
\right.
\label{bcPl}
\end{eqnarray}
Here we are assuming equal 5D gauge couplings on the two sides for simplicity.
In case of different $g_5$'s, Eqs.~(\ref{bcPl}) will be modified according to the rules given in the previous section. 
However, this restricted set of parameters  is sufficient for our purposes, because, as we will comment later, in the gauge sector the effect of different 5D couplings is equivalent to different AdS curvature scales on the two sides. 

Regarding the IR breaking, we will study three limits: that in which it comes from a
Higgs on both IR branes (``Higgs---top-Higgs"), that in which most electroweak
symmetry breaking is from a higgsless breaking while the top gets its mass
from a Higgs (``higgsless---top-Higgs"), and that in which we have higgsless
boundary conditions on both IR branes (``higgsless---higgsless").

For future convenience, we first recall the expressions for $M_W$ in the Randall-Sundrum scenario with an IR brane Higgs \cite{RS,EffRS} and in the higgsless case \cite{CGPT}, in the limit $g_{5L} = g_{5R}$ at leading log order:
\beq
M_{W;Higgs}^{2} &\approx& \frac{g_5^2}{R\log \frac{R^\prime}{R}} \frac{R^2 v^2}{4 R^{\prime 2}}\,, \label{eq:MWhiggs}\\
M_{W;higgsless}^{2} &\approx& \frac{1}{R^{\prime 2} \log \frac{R^\prime}{R}}\,.\label{eq:MWhiggsless}
\eeq
Also, independent of the IR brane boundary conditions, we can express the 4D gauge couplings in terms of the 5D parameters:
\begin{eqnarray}
\frac{1}{e^2} &=& \left( \frac{1}{g_{5L}^2} + \frac{1}{g_{5R}^2} + \frac{1}{\tilde g_5^2} \right) \left(R_w \log \frac{R_w^\prime}{R_w} + R_t \log \frac{R_t^\prime}{R_t} \right)\,,\\
\tan \theta_W^2 &=& \frac{g_{5R}^2 \tilde g_5^2}{g_{5L}^2 (g_{5R}^2 + \tilde g_5^2)}\,,\\
g^2 &=& \frac{g_{5L}^2}{R_w \log \frac{R_w^\prime}{R_w} +  R_t \log \frac{R_t^\prime}{R_t}}\,.
\end{eqnarray}
Note that the the 5D gauge couplings are related to the 4D ones via the total volume of the space, namely the sum of the two AdS spaces.
Moreover, as we have two sets of ``Planck'' and ``TeV'' scales, we are assuming the logs to be of the same order in the expansion.

\subsection{Higgs---top-Higgs}

We first assume that on each IR brane we have a scalar Higgs field, transforming as a bifundamental under $SU(2)_L\times SU(2)_R$.
They develop VEVs $v_w$ and $v_t$ that break this group to $SU(2)_D$, leaving $U(1)_{B-L}$ unbroken.
We also take the localized Lagrangians in the form (\ref{eq:Llocgen}), so that $v_w \sim 1/R_w$ and $v_t \sim 1/R_t$. 
The VEVs will generate a mass term for the combination $A^L_\mu - A^R_\mu$, while the fields related to the unbroken subgroup
\begin{equation}
A_\mu^{Da} = \frac{g_{5L}^2 g_{5R}^2}{g_{5L}^2 + g_{5R}^2} \left(\sfrac{1}{g_{5L}^2} A_\mu^{La} + \sfrac{1}{g_{5R}^2} A_\mu^{Ra} \right)\,,
\end{equation}
and $B_\mu$ have Neumann BCs.
The complete set of BCs is:
\beq
&
{\rm at }\  z=R_w^\prime
&
\left\{
\begin{array}{l}
\partial_z (\sfrac{1}{g_{5L}^2} A^{L\,a}_\mu + \sfrac{1}{g_{5R}^2} A^{R\,a}_\mu)  = 0\,, \\
\tv{15}
\partial_{z} (A^{L\,a}_\mu - A^{R\,a}_\mu) = - \sfrac{v_{w}^2}{4} \frac{R_w}{R_w^\prime} \left(g_{5L}^2 + g_{5R}^2\right) (A^{L\,a}_\mu  - A^{R\,a}_\mu)\,, \\
\tv{15}
\partial_z B_\mu   = 0\,,
\end{array}
\right.
\label{BCHiggs}
\eeq
and similarly on the other IR brane.

To determine the spectrum in this case, we use the expansion of the Bessel function for small argument (assuming $M_A R^\prime \ll 1$),
\beq
\psi^{(A)}(z) \approx c^{(A)}_0 + M_A^2 z^2 \left( c^{(A)}_1 - \frac{c^{(A)}_0}{2} \log \frac{z}{R} \right)
\eeq
and solve (much as in \cite{CGPT}). We assume that $v_i R_i$ is small. We find for the $W$ mass:
\beq
M_W^2 &\approx& \frac{g_{5L}^2}{R_w \log \frac{R_w^\prime}{R_w} + R_t \log \frac{R_t^\prime}{R_t} } \left( \left( \frac{R_w}{R_w^\prime}\right)^2 \frac{v_w^2}{4} + \left(\frac{R_t}{R_t^\prime}\right)^2 \frac{v_t^2}{4} \right) \nonumber \\
&=& \frac{g^2}{4} \left( \left( \frac{R_w}{R_w^\prime}\right)^2 v_w^2 + \left(\frac{R_t}{R_t^\prime}\right)^2 v_t^2 \right)\,.
\label{WMassHiggsHiggs}
\eeq
This is of the form one would expect for a gauge boson obtaining its mass from two Higgs bosons. Note that the natural scale of the two contributions is $\sfrac{1}{R_w^{\prime 2}}$ and $\sfrac{1}{R_t^{\prime 2}}$ respectively.

We will not discuss this case at any length, as viable Randall-Sundrum models with Higgs boson exist~\cite{ADMS}. 
We simply note that one can construct a variety of models analogous to two-Higgs doublet models, in which one has distinct KK spectra for particles coupling to different Higgs bosons.

\subsection{Higgsless---top-Higgs}

Next we consider a case in which the IR brane at $R_w^\prime$ has a higgsless boundary condition and is responsible for most of the electroweak symmetry breaking, while a top-Higgs on the brane at $R_t^\prime$ makes some smaller contribution to electroweak symmetry breaking. In this case we have distinct BC's:
\beq
&
{\rm at }\  
z=R_w^\prime
&
\left\{
\begin{array}{l}
\partial_z (\sfrac{1}{g^2_{5L}} A^{L\,a}_\mu +
\sfrac{1}{g^2_{5R}} A^{R\,a}_\mu)  = 0\,, \\
\tv{15}
A^{L\,a}_\mu  - A^{R\,a}_\mu =0, \
\partial_z B_\mu   = 0\,,
\end{array}
\right.
\label{BChiggsless}
\eeq
while at $z=R_t^\prime$ we have the same BCs as in Eq.~(\ref{BCHiggs}).

Solving as above, we find:
\beq
M_W^2 &\approx& \frac{g_{5L}^2}{R_w \log \frac{R_w^\prime}{R_w} + R_t \log \frac{R_t^\prime}{R_t} } \left( \frac{R_w}{R_w^{\prime 2}} \frac{2}{g_{5L}^2+g_{5R}^2} +  \left(\frac{R_t}{R_t^\prime}\right)^2 \frac{v_t^2}{4} \right) \nonumber \\
&=& g^2 \left(\frac{2 R_w}{g_{5L}^2+g_{5R}^2} \frac{1}{R_w^{\prime 2}}+ \left(\frac{R_t}{R_t^\prime}\right)^2 \frac{v_t^2}{4} \right)\,.
\label{WMasshiggslessHiggs}
\eeq
This again takes the form of a sum of squares, with one term of the form found in the usual higgsless models (\ref{eq:MWhiggsless}) and one term in the form of an ordinary contribution from a Higgs VEV (\ref{eq:MWhiggs}). 
Our boundary condition is a limit as $v_{w} \rightarrow \infty$ of the previous case, so we expect that for intermediate values of $v_{w}$, its contribution will level off smoothly to a constant value. 

If we want to disentangle the top mass from the electroweak symmetry breaking sector, we assume that $v_t R_t \ll 1$ so that the $W$mass comes mostly from the higgsless AdS.
In this case the perturbative unitarity in longitudinal $W$scattering will be restored by the gauge boson resonances.
The physical top-Higgs, arising from the AdS$_t$, will not contribute much to it and will mostly couple to the top quark.
We will come back to the physics of the third generation in Section~\ref{sec:thirdgen}.

\subsection{Higgsless---higgsless}

Finally, we consider a case in which both IR branes have higgsless boundary conditions Eq~(\ref{BChiggsless}).
In this case, as expected, we find:
\beq
M_W^2 &\approx& \frac{g_{5L}^2}{R_w \log \frac{R_w^\prime}{R_w} + R_t \log \frac{R_t^\prime}{R_t} }\frac{2}{g_{5L}^2+g_{5R}^2} \left( \frac{R_w}{R_w^{\prime 2}} +\frac{R_t}{R_t^{\prime 2}}\right) \nonumber \\
&=& g^2 \left(\frac{2 R_w}{g_{5L}^2+g_{5R}^2} \frac{1}{R_w^{\prime 2}}+ \frac{2 R_t}{g_{5L}^2+g_{5R}^2} \frac{1}{R_t^{\prime 2}} \right),
\label{WMasshiggslesshiggsless}
\eeq
where we have grouped the terms for later convenience in discussing the holographic interpretation. Note that in the symmetric limit $R_w=R_t=R$, $R_t^\prime=R_w^\prime=R^\prime$, we recover the usual (one bulk) higgsless result (\ref{eq:MWhiggsless}). Our expression can be reformulated in another useful way:
\begin{equation}
M_W^2 =\frac{2 g_{5L}^2}{g_{5L}^2+g_{5R}^2} \left( \frac{1}{R_w^{\prime 2}}  + \frac{R_t}{R_w} \frac{1}{R_t^{\prime 2}}  \right) \frac{1}{\log \frac{R_w^\prime}{R_w} + \frac{R_t}{R_w} \log \frac{R_t^\prime}{R_t}}\,,
\end{equation}
In this formulation there is a manifest limit where the contribution of the AdS$_t$ is small, namely if the volume of the new space is smaller that the volume of the old one: $R_t \ll R_w$.
In this case the contribution of the top-sector to $M_W$ is suppressed.
It is interesting to note that this property is actually related to the relative size of the 5D gauge coupling and the warping factor.
  From the matching conditions, we find that $g_5^2$ is of order to the total volume of the space.
The limit we are interested in is in fact when the 5D gauge coupling is larger than the warp factor in the second AdS, namely $g_5^2 \approx R_w \gg R_t$.
On the other hand, if we assume that there also are different gauge couplings on the two spaces, each one of the order of the local curvature, the decoupling effect disappears.
In this sense, at the level of the gauge bosons, a hierarchy between the curvatures is equivalent to a hierarchy between the bulk gauge couplings.

It is also interesting to study the spectrum of the resonances: at leading order in the log-expansion, they decouple into two towers of states proportional to the two IR scales. Namely:
\begin{equation} \label{eq:KK0}
M_{w^\prime}^{(n)} \approx \frac{\mu^{(n)}_{0, 1}}{R_w^\prime}\,, \quad M_{t^\prime}^{(n)} \approx \frac{\mu^{(n)}_{0, 1}}{R_t^\prime}\,,
\end{equation}
where the numbers $\mu^{(n)}_{0, 1}$ are respectively the zeros of the Bessel functions $J_0 (x)$ and $J_1 (x)$.
This is true irrespective of the BC's on the TeV brane.
In the higgsless case, with $R_t \ll R_w$, the tower of states proportional to $1/R_w^\prime$ receives corrections suppressed by a log:
\begin{equation} \label{eq:KK1} \begin{array}{rcl}
M_{(0)}^{(n)} &\approx &\frac{1}{R_w^\prime} \left( \mu^{(n)}_0 + \frac{\pi}{2} \frac{g_{5R}^2}{g_{5L}^2 + g_{5R}^2} \frac{Y_0(\mu^{(n)}_0)}{J_1(\mu^{(n)}_0)} \frac{1}{\log \frac{R_w^\prime}{R_w}} + \mathcal{O} (\log^{-2}) \right)\,,\\
M_{(1)}^{(n)} &\approx &\frac{1}{R_w^\prime} \left( \mu^{(n)}_1 + \frac{\pi}{2} \frac{g_{5L}^2}{g_{5L}^2 + g_{5R}^2} \frac{Y_1(\mu^{(n)}_1)}{J_2(\mu^{(n)}_1)-J_0(\mu^{(n)}_1)} \frac{1}{\log \frac{R_w^\prime}{R_w}} + \mathcal{O} (\log^{-2}) \right)\,.
\end{array} \end{equation}
This is equivalent to the states of a one brane model, up to corrections suppressed by $R_t/R_w$.
On the other hand, the corrections to the tower proportional to $1/R_t^\prime$ are always suppressed by $R_t/R_w$.

Finally, we can compute the oblique observables.
Due to the custodial symmetry, we find $T \approx 0$, and for the case when the light fermions are localized close to 
the Planck brane the $S$-parameter is:
\begin{equation}
S \approx \frac{6 \pi}{g^2} \frac{2 g_{5L}^2}{g_{5L}^2 + g_{5R}^2} \frac{R_w + R_t}{ R_w \log \frac{R_w^\prime}{R_w} + R_t \log \frac{R_t^\prime}{R_t}}\,.
\end{equation}
Also in this case, the contribution from the additional AdS$_t$ is suppressed by the ratio $R_t/R_w$. Just as for the simple 
higgsless case the contribution to the $S$-parameter can be suppressed by moving the light fermions into the bulk and thus
reducing their couplings to the KK gauge bosons~\cite{CuringIlls,MSUdeloc}.

\section{The CFT Interpretation}
\label{sec:CFT}
\setcounter{equation}{0}
\setcounter{footnote}{0}

We would now like to interpret the formulae in the previous section in the 4D CFT language. Since we now have
two bulks and two IR branes, it is natural to assume that there would be two separate CFT's corresponding to 
this system. Each of these CFT's has its own set of global symmetries, given by the gauge fields in each of the 
bulks. The gauge fields which vanish on the Planck brane correspond to genuine global symmetries, however the ones that 
are allowed to propagate through the UV brane will be weakly gauged. Since there is only one set of light (massless) 
modes for these fields, clearly only the diagonal subgroup of the two independent global symmetries of the 
two CFT's will be gauged. 

The first test for this interpretation is in the spectrum of the KK modes.
Indeed, in the limit  where we remove the Planck brane, the only object that links the two AdS spaces, we should find two independent towers of states, each one given by the bound states of the two CFT's and with masses proportional to the two IR scales.
This is exactly the structure we found in Eq.~(\ref{eq:KK0}).
Moreover, the $W$boson gets its mass via a mixing with the tower of KK modes, so that it can be interpreted as a mixture of the elementary field and the bound states.
This mixing also introduces corrections to the simple spectrum described above, suppressed by the log.
In the limit $R_t \ll R_w$ the $W$mass comes from the AdS$_w$ side, so that we expect large corrections to the states with mass proportional to $1/R_w^\prime$ and small corrections to the states with mass $\sim 1/R_t^\prime$.
This is confirmed by Eqs.~(\ref{eq:KK1}).

Let us now discuss in detail the interpretation of the electroweak symmetry breaking mechanisms described in 
the previous section. We know that the interpretation of a IR brane Higgs field is that the CFT is forming a 
composite scalar bound state which then triggers electroweak symmetry breaking, while the interpretation of the higgsless boundary 
conditions is that the  CFT forms a condensate that gives rise 
to electroweak symmetry breaking (but no composite scalar). Thus these latter models can be viewed as extra dimensional
duals of technicolor type models. What happens when we have the setup with two IR branes? Each of the 
two CFT's will break its own global $SU(2)_L\times SU(2)_R$ symmetries to the diagonal subgroup either via
the composite Higgs or via the condensate. We can easily test these conjectures by deriving, just based on this correspondence, the formulae obtained in 
the previous section via explicitly solving the 5D equations of motion. 

In order to be able to do that we need to find the explicit expression for the pion decay constant $f_\pi$ of 
these CFT's. This can be most easily found by comparing the generic expression of the $W$-mass in higgsless models
\begin{equation}
M_W^2=\frac{2 g_{5L}^2}{g_{5L}^2+g_{5R}^2} \frac{1}{R'^2 \log \frac{R'}{R}}
\end{equation}
with the expression for the $W$-mass in a generic technicolor model with a single condensate
\begin{equation}
M_W^2 =  g^2 f_\pi^2 \ .
\end{equation}
Using the tree-level relation between $g$ and $g_{5L}$ in RS-type models we find that 
\begin{equation}
f_\pi^2 = \frac{2R}{g_{5L}^2+g_{5R}^2} \frac{1}{R'^2}.
\end{equation}
The usual interpretation of this formula~\cite{BN} 
in terms of large-$N$ QCD theories is by comparing it to the relation
\begin{equation}
f_{\pi} \sim \frac{\sqrt{N}}{4\pi} m_\rho ,
\end{equation}
where $m_\rho$ is the characteristic mass of the techni-hadrons, and $N$ is the number of colors.
In our case $m_\rho \sim 1/R'$, so the number of colors would be given by
\begin{equation}
N \sim \frac{32 \pi^2 R}{g_{5L}^2+g_{5R}^2}.
\end{equation}
Note that one will start deviating from the large N limit once $\frac{R}{g_{5L}^2+g_{5R}^2} \ll 1$.

For the case when there is a composite Higgs the effective VEV
(as always in RS-type models) is nothing but the warped-down version
of the Higgs VEV
\begin{equation}
v_{{\rm \it eff}} =v \frac{R}{R'}\ .
\end{equation}

Now we can use this formula to derive expressions for the $W$and $Z$ masses in the general cases with two IR branes. Based
on our correspondence both of the CFT's break the gauge symmetry, either via a composite Higgs or via 
the condensate. Since the gauge group is the diagonal subgroup of the two global symmetries, we simply need
to add up the contribution of the two CFT's. So for the Higgs---top-Higgs case we would expect
\begin{equation}
M_W^2= \frac{g^2}{4} (v_{{\rm \it eff},w}^2+v_{{\rm \it eff},t}^2)=
\frac{g^2}{4} \left( \left( \frac{R_w}{R_w^\prime}\right)^2 v_w^2 + \left(\frac{R_t}{R_t^\prime}\right)^2 v_t^2 \right)\,,
\end{equation}
which is in agreement with (\ref{WMassHiggsHiggs}). In the mixed higgsless---top-Higgs case we expect the 
$W$mass to be given by
\begin{equation}
M_W^2= g^2 (f_{\pi, w}^2 + \frac{v_{{\rm \it eff},t}^2}{4}) = g^2 \left(\frac{2 R_w}{g_{5L}^2+g_{5R}^2} \frac{1}{R_w^{\prime 2}}+ \left(\frac{R_t}{R_t^\prime}\right)^2 \frac{v_t^2}{4} \right)\,,
\end{equation}
which is again in agreement with Eq. (\ref{WMasshiggslessHiggs}). Finally, in the higgsless---higgsless case we
expect the $W$ mass  to be given by
\begin{equation}
M_W^2 = g^2 (f_{\pi, w}^2 + f_{\pi, t}^2) = g^2 \left(\frac{2 R_w}{g_{5L}^2+g_{5R}^2} \frac{1}{R_w^{\prime 2}}+ \frac{2 R_t}{g_{5L}^2+g_{5R}^2} \frac{1}{R_t^{\prime 2}} \right)\,,
\label{eq:WMHless2}
\end{equation}
again in agreement with the result of the explicit calculation (\ref{WMasshiggslesshiggsless}).

\section{Top-pions}
\label{sec:toppion}
\setcounter{equation}{0}
\setcounter{footnote}{0}

\subsection{Top-pions from the CFT correspondence}

We can see from the match of the expressions of the $W$ masses above that the CFT picture is reliable. However, the CFT
picture has one additional very important prediction for this model: the existence of light pseudo-Goldstone
bosons, which are usually referred to as top-pions in the topcolor literature. The emergence of these can be easily 
seen from the gauge and global symmetry breaking structure. We have seen that there are two separate CFT's, 
each of which has its own $SU(2)_L\times SU(2)_R$ global symmetry. Only the diagonal $SU(2)_L\times U(1)_Y$ is 
gauged. Both CFT's will break their respective global symmetries as $SU(2)_L\times SU(2)_R\to SU(2)_D$. Thus
both CFT's will produce three Goldstone bosons, while the gauge symmetry breaking pattern is the usual one for
the SM $SU(2)_L\times U(1)_Y\to U(1)_{QED}$, so only three gauge bosons can eat Goldstone bosons. Thus we will 
be left with three uneaten Goldstone modes, which will manifest themselves as light (compared to the resonances) 
isotriplet scalars. They will not be exactly massless, since the fact that only the diagonal $SU(2)_L\times U(1)_Y$
subgroup is gauged will explicitly break the full set of two $SU(2)_L\times SU(2)_R$ global symmetries.
Thus we expect these top-pions to obtain mass from one-loop electroweak interactions. We can give a rough estimate 
for the loop-induced size of the top-pion mass. For this we need to know which linear combination of the two 
Goldstone modes arising from the two CFT's will be eaten. This is dictated by the Higgs mechanism, and the usual
expression for the uneaten Goldstone boson is
\begin{equation}
\Phi_{top \pi}=\frac{f_{\pi, t} \Phi_w - f_{\pi, w} \Phi_t}{\sqrt{f_{\pi, w}^2 + f_{\pi, t}^2}}\ ,
\end{equation}
where $\Phi_{w,t}$ are the isotriplet Goldstone modes from the two CFTs, and the $f_\pi$'s should be substituted
by $f_{\pi, i}\to \sfrac{1}{2} v_{{\rm \it eff},i}$ ($i=w,t$) in case we are not in the higgsless limit. Since the scale of the resonances in the two
CFT's are given by $m_{\rho, i}=\frac{1}{R'_{i}}$, we can estimate the loop corrections to the top pion mass to be
of order 
\begin{equation}
m_{top \pi}^2 \sim \frac{g^2}{16\pi^2 (f_{\pi, w}^2 + f_{\pi, t}^2)} \left( \frac{f_{\pi, w}^2}{R_w^{\prime 2}} + \frac{f_{\pi, t}^2}{R_t^{\prime 2}}
\right).
\end{equation}
Experimentally, these top-pions should be heavier than $\sim $100 GeV. 
We will provide a more detailed expression for their masses when we discuss them in the 5D picture.

We can also estimate the coupling of these top-pions to the top and bottom quarks, 
assuming that the top pion lives mostly in the 
CFT that will give a rather small contribution to the $W$mass, but a large contribution to the top mass.
The usual CFT interpretation of the top and bottom mass is the following~\cite{holoFermions}: the left-handed top and bottom are elementary fields living in a doublet of the $(t_L,b_L)$.
The right handed fields have a different nature: the top is a composite massless mode contained in a doublet under the local $SU(2)_R$, while the bottom is an elementary field weakly mixed with the CFT states to justify the lightness of the bottom. 
Assuming that a non-linear sigma model is a good description for the top-pions, we find that 
the top and bottom masses can be written in the following $SU(2)_L\times SU(2)_R$ invariant form:
\begin{equation}
(\bar{t}_R,\bar{b}_R/N_R^b ) U^\dagger_R\left( \begin{array}{cc} m_t \\ & m_t \end{array} \right) U_L \left( \begin{array}{c}
t_L \\ b_L \end{array} \right) +h.c.
\end{equation}
Here the suppression factor $N_R^b$ is due to the fact that only a small mixture of the right handed bottom is
actually composite. To obtain the correct masses we will need $N_R^b\sim m_b/m_t$. 
If the top-pion is mostly the Goldstone boson from the CFT$_t$ that gives the top mass then
$U_L = U^\dagger_R \sim e^{i \Phi^a \tau^a/2 f_{\pi t}}$, which implies that the coupling of the top-pion to the 
top-bottom quarks will be of the form
\begin{equation}
\frac{m_{t}}{2 f_{\pi, t}} (t_L \Phi^3 \bar{t}_R + \sqrt{2} b_L \Phi^- \bar{t}_R) + \frac{m_{b}}{2 f_{\pi, t}} 
(b_L \Phi^3 \bar{b}_R + \sqrt{2} t_L \Phi^- \bar{b}_R) + h.c.\,.
\end{equation}
Thus we can see that the 
couplings involving $t_R$ are proportional to $m_{t}/f_{\pi, t}$, which will be large in the limit when the CFT does not 
contribute significantly to the $W$mass. A similar argument can be made in the limit when the electroweak 
symmetry breaking in CFT$_t$ appears mostly from the VEV of a composite Higgs. In this case this Higgs VEV has to 
produce the top mass, and so we can show that the couplings of the top pions will be of order
$m_{t}/v_{eff,t}$. Thus these couplings will be unavoidably large both in the higgsless and the higgs limit of the
CFT$_t$, and one has to worry whether these couplings would induce additional shifts in the value of the 
top quark mass and the $Zb\bar{b}$ couplings.

\subsection{Properties of the top-pion from the 5D picture.}

We have shown in Section~\ref{sec:warmup} how such massless modes appear in the 5D picture from the modified BC of the $A_5$ fields.
We would like now to study in more detail their properties, already inferred from the CFT picture.
The first check is to show how the strong coupling with the top and bottom arises in the 5D picture.
Let us recall how the fermion masses are generated through brane localized interactions~\cite{BPR,CGHST}.
The left- and right-handed fermions are organized in bulk doublets of $SU(2)_L$ and $SU(2)_R$ respectively, where specific boundary conditions are picked  in order to leave chiral zero modes, and the localization of the zero modes is controlled by two bulk masses $c_L$ and $c_R$ in units of $1/R$.
The Higgs localized on the IR brane allows one to write a Yukawa coupling  linking the L and R doublets (in the higgsless limit this corresponds to a Dirac mass term): this gives a common mass to the up- and down- type quarks, due to the unbroken $SU(2)_D$ symmetry.
The mass splitting can be then recovered adding a large kinetic term localized on the Planck brane for the $SU(2)_R$ component of the lighter quark~\cite{BPR}.
A similar mechanism can be used to generate lepton masses.

In the higgsless---top-Higgs scenario, the localized Yukawa couplings can be written as:
\begin{equation}
\int d z \left( \frac{R_t}{R_t^\prime} \right)^4 \delta (z-R_t^\prime)\, \lambda_{top} R_t\, \left( \chi_L \phi \eta_R + h.c. \right)\,,
\end{equation}
where $\lambda_{top}$ is a dimensionless quantity, and the 5D fields $\chi_L$ ($\eta_R$) are the left- (right-) handed components that contain the top-bottom zero modes.
Expanding the Higgs around the VEV:
\begin{equation}
\phi = \frac{v_t}{\sqrt{2}} \left( 1+\frac{R_t^\prime}{R_t} \frac{h_t + i \pi_t^a \sigma^a}{v_t} \right)\,,
\end{equation}
where the warp factor takes into account the normalization of the scalars, we can find the trilinear interactions involving the top-pion triplet $\pi_t^a$ and the top-Higgs $h_t$.
In the following we will assume that the fermion wave functions are given by the zero modes, basically neglecting the backreaction of the localized terms: this approximation is valid as long as the top mass is small with respect to the IR brane scale, namely $m_t R_t^\prime < 1$.
The wave functions are then
\begin{equation}
\chi_L = \frac{1}{\sqrt{R_t}} \left( \frac{z}{R_t} \right)^{2-c_L} \left( \begin{array}{c}
t_l/N_L^t\\
b_l/N_L^b
\end{array} \right)\,, \quad 
\eta_R = \frac{1}{\sqrt{R_t}} \left( \frac{z}{R_t} \right)^{2+c_R} \left( \begin{array}{c}
t_r/N_R^t\\
b_r/N_R^b
\end{array} \right)\,.
\end{equation}
The normalizations for the L-fields are given by the bulk integral of the kinetic term, and are the same for top and bottom $N_L^t = N_L^b$.
On the other hand, for the $b_r$ field, the wave function is dominated by the localized kinetic term on the Planck brane.
This is the source for the splitting between the top and bottom mass, so that
\begin{equation}
\frac{N_R^b}{N_R^t} = \frac{m_t}{m_b}\,.
\end{equation}

With this in mind, we find that:
\begin{equation}
m_t = \frac{\lambda_{top} v_t R_t}{\sqrt{2}R_t^\prime} \left( \frac{R_t^\prime}{R_t} \right)^{1+c_R-c_L} \frac{1}{N_L N_R^t}\,,
\end{equation}
and the couplings can be written as
\begin{equation}
\frac{m_t}{v_{{\rm \it eff},t}} \left( t_r, \frac{m_b}{m_t} b_r \right) \left( h_t + i \sigma^a \pi_t^a \right) \left( \begin{array}{c}
t_l\\ b_l \end{array} \right)\,. 
\end{equation}
It is clear that the $t_l t_r$ and $b_l t_r$ couplings are enhanced by the ratio $m_t/v_{{\rm \it eff},t}$, while the couplings involving the r-handed $b$ will be suppressed by the bottom mass.
This will lead to possibly large and incalculable contributions to the top-mass and the $b_l$ coupling with the $Z$, however it is still plausible to have a heavy top in such models.

In the higgsless---higgsless limit the situation is more complicated: the top-pion is a massless mode of the $A_5$ of the broken generators of $SU(2)_L\times  SU(2)_R$ (no Higgs is present in this  limit) and its couplings are determined by the gauge interactions in the bulk, thus involving non trivial integrals of wave functions.
However, as suggested by the CFT interpretation, we will get similar couplings, with $v_{{\rm \it eff},t}$ replaced by $f_{\pi, t}$, and again the requirement that $f_{\pi, t}\ll M_W$ will introduce strong coupling.
This is a generic outcome from the requirement that the symmetry breaking scale that gives rise to the top mass is smaller than the electroweak scale.

An important, and in some sense related issue is the mass of the top-pion.
As already mentioned, it will pick up a mass at loop level, generated by the interactions that break the two separate global symmetries.
The top only couples to one CFT, so that its interactions cannot contribute to the top-pion mass.
The net effect, although non-calculable due to strong coupling, is to renormalize the potential for the 
Higgs localized on the new IR brane.
On the other hand, the only interactions that break the global symmetries are the gauge interactions that can propagate from one boundary to the other.
In the higgsless case, assuming weak coupling, we expect a contribution of the form:
\beq \label{eq:mA5}
m^2_{A_5} = \frac{C(r)}{\pi} \frac{g_{5L}^2 + g_{5R}^2}{R_t}\frac{1}{R_t^{\prime 2}} F(R_t/R_t^\prime),
\eeq
where $F(R_t/R_t^\prime)$ is typically order 1~\cite{CNP}.
In the phenomenologically interesting region, this effect is not calculable, as the gauge KK modes are strongly coupled.
However we still expect the mass scale to be set by $1/R_t^\prime$.
In the case of the top-Higgs, the gauge KK modes are weakly coupled to the localized Higgses, so a loop expansion makes sense and Eq.~(\ref{eq:mA5}) is a good estimate of the mass.

Another interesting issue is the sign of the mass squared.
Indeed, a negative mass square for the charged top-pion would signal a breakdown of the electromagnetic U(1)$_{QED}$, and would generate a mass for the photon.
The gauge contribution is usually expected to be positive.
In other words, we need to make sure that our symmetry breaking pattern is stable under radiative corrections.
A useful way to think about it is the following: from the effective theory point of view, we have a two Higgs model.
The tree level potential consists of two different and disconnected potentials for the two Higgses, so that two different 
$SU(2)_L\times SU(2)_R$ global symmetries can be defined.
After the Higgses develop VEVs, we can use a gauge transformation to rotate away the phase of one of them, but a relative phase could be left.
In other words the tree-level potential itself does not guarantee that the two VEVs are aligned and a $U(1)$ is left unbroken.
Once we include the radiative contribution to the potential, the qualitative discussion does not change: some mixing terms will be generated by the gauge interactions, lifting the massless pseudo-Goldstone bosons, but in general a relative phase could be still present.
So, we need to assume that the two VEVs are aligned, maybe by some physics in the UV. There is an analogous vacuum alignment problem that  arises in the SM if we consider the limit of small $u$ and $d$ masses, where the dominant contribution to the mass of the $\pi^\pm$ comes from a photon loop.  In that case, QCD spectral density sum rules can be used to show that $m_{\pi^\pm} >0$ \cite{peskinpresskill}. Thus in the higgsless---higgsless case  it is possible that the dual CFT dynamics can ensure the correct vacuum alignment, as happens in QCD-like technicolor theories \cite{peskinpresskill}.

A possible extension of the model is considering a bulk Higgs instead of brane localized Higgses, in order to give an explicit mass to the top-pion in analogy with the QCD case studied in~\cite{AdSQCD}.
We imagine that we have a single Higgs stretching over the two bulks.
The generic profile for the Higgs VEV along the AdS space will be~\cite{bulkHiggs}:
\begin{equation}
v (z) \sim a \left( \frac{z}{R} \right)^{\Delta_+} + b \left( \frac{z}{R} \right)^{\Delta_-}\,,
\end{equation}
where the exponents $\Delta_{\pm} = 2 \pm \sqrt{4+M^2_{bulk} R^2}$ are determined by the bulk mass of the scalar.
The bulk mass controls the localization of the VEV near the two IR branes, and in the large mass limit we recover the two Higgses case: all the resonances become very heavy and decouple, except for one triplet that becomes light and corresponds to the top-pion.
Indeed, its mass will be proportional to the value of the VEV on the Planck brane, that is breaking the two global symmetries explicitly.
In the CFT picture, the bulk VEV is an operator that connects the two CFTs and gives a tree level mass to the top-pion.
However, the bulk tail will also contribute to the $W$mass: we numerically checked in a simple case that in any interesting limit, when the bulk Higgs does not contribute to unitarity, the tree level mass is negligibly small.
Nevertheless, this picture solves the photon mass issue: indeed we have only one Higgs.
In other words, the connection on the Planck brane is forcing the two VEVs on the boundaries to be aligned.

To summarize, the top-pion will certainly get a mass at loop level, whose order of magnitude can be estimated to be at least one 
loop factor times $1/R_t^\prime$.
Moreover, a tiny explicit mixing between the two CFT, induced by a connections of the two VEVs on the Planck brane, would be 
enough to stabilize the symmetry breaking pattern and preserve the photon from getting a mass. 
It would be interesting to analyze more quantitatively these issues which we leave for further studies.

\section{Phenomenology of the Two IR Brane Models}
\label{sec:thirdgen}
\setcounter{equation}{0}
\setcounter{footnote}{0}

\subsection{Overview of the various models}

The main problematic aspect of higgsless models of electroweak symmetry breaking is the successful incorporation
of a heavy top quark into the model without significantly deviating from the measured values of the $Zb_l\bar{b}_l$ 
coupling~\cite{CuringIlls}.
The reason behind this tension is that there is an upper bound on the mass of a fermion localized at least partly on the Planck brane given by
\begin{equation}
m_f^2 \leq \frac{2}{R'^2 \log \frac{R'}{R}} \ .
\label{eq:topbound}
\end{equation}
Since in the case of a single TeV brane the value of $R'^2 \log \frac{R'}{R}$ is determined by the $W$ mass, the only 
way to overcome this bound is by localizing the third generation quarks on the TeV brane. However the region around the 
TeV brane is exactly the place where the wave functions of the $W$and $Z$ bosons are significantly modified, thus leading
to large corrections in the $Zb\bar{b}$ coupling. 

The main motivation for considering the setups with two IR branes is to be able to separate the mass scales 
responsible for the generation of the $W$ mass  and the top mass. Thus as discussed before, we are imagining a setup
where electroweak symmetry breaking is coming dominantly from a higgsless model like the one discussed in~\cite{CuringIlls},
while the new side is responsible for generating a heavy top quark. The gauge bosons would of course have to live in both
sides, while of the fermions only the third generation quarks would be in the new side. We have seen that in the 
higgsless---higgsless limit the $W$ mass  is given by (\ref{eq:WMHless2})
\begin{equation}
M_W^2 =g^2 (f_{\pi, w}^2 + f_{\pi, t}^2)=\left( \frac{1}{R_w^{\prime 2}}  + \frac{R_t}{R_w} \frac{1}{R_t^{\prime 2}}  \right) \frac{1}{\log \frac{R_w^\prime}{R_w} + \frac{R_t}{R_w} \log \frac{R_t^\prime}{R_t}}\,.
\label{Wmasssimple}
\end{equation}
In order to ensure that the dominant contribution to the $W$mass arises from a higgsless model as in \cite{CuringIlls} we need to 
suppress the contribution of the new side by choosing $\frac{R_t}{R_w}\ll 1$, which in the CFT picture corresponds to
$f^2_{\pi, t} \ll f^2_{\pi, w}$. This way one can choose parameters on the ``old side'' similarly as in the usual
higgsless models, that is $1/R_w^\prime \sim 300$ GeV, $\log \frac{R_w^\prime}{R_w} \sim 10$, resulting in a KK modes of the $W$ 
and $Z$ of about 700 GeV. The couplings will be slightly altered from the one-bulk higgsless model, but will remain close enough 
that we can maintain perturbative unitarity up to scales of about 10 TeV, provided the new side contributes only about 1\% of 
the $W$ mass. We will substantiate this claim numerically in a later section. All this can be achieved independently of the 
choice of $R_t^{\prime -1}$! Thus we can still make $1/R_t^\prime$ quite a bit bigger than the TeV scale on the old side $1/R_w^\prime$,
making it possible to obtain a large top quark mass by circumventing the bound (\ref{eq:topbound}).

However, this framework is not without potential problems: as discussed in the previous section there is a light top-pion 
pseudo-Goldstone mode with a generically  large 
coupling to the top quark, irrespective of the value of the VEV of the Higgs on the TeV brane. 
A more worrisome problem arises from the limit $f^2_{\pi, t} \ll f^2_{\pi, w}$. In 
general, a condition for a trustworthy 5D effective field theory is that $b_{CFT} = 8 \pi^2 \sfrac{R}{g_5^2} \gg 1$ \cite{APR}, 
a condition that will be violated in the new bulk when $f^{2}_{\pi, t}$ is too small. Violation of this condition will result
in large 4D couplings among the KK modes of the AdS$_t$, whose masses are proportional to $1/R_t^\prime$, which could potentially give rise to large incalculable shifts in the 
expected values of the masses and couplings of these modes. Note however, that the coupling of the light $W$and $Z$ will
never be strong to these KK modes: gauge invariance will make sure that the couplings of the $W$and $Z$ are of the order $g$ to 
all of these modes. Thus the expressions for the electroweak precision observables 
will be shielded by at least an electroweak loop suppression from  the potentially strong couplings of the KK modes.
Moreover, the unitarity in the longitudinal $W$scattering is still maintained by the KK modes of the AdS$_w$, which are weakly coupled with the KK modes of the new AdS.
This is again true as long as the $W$mass mostly comes from the old side.

Thus there is a tension in the 
higgsless---higgsless case between constraints from perturbative unitarity of $WW$ scattering 
which will want $f^2_{\pi, t} \ll f^2_{\pi, w}$, and constraints from 5D effective field 
theory. Of course, we could evade the perturbative unitarity problem by lowering $R_t^{\prime -1}$ to near the scale $R_w^{\prime -1}$, 
but then the new side would merely be a copy of the old side and one would again start running into trouble with the 
third generation physics. 
\begin{figure}[tb]
\begin{center}
\includegraphics[width=14cm]{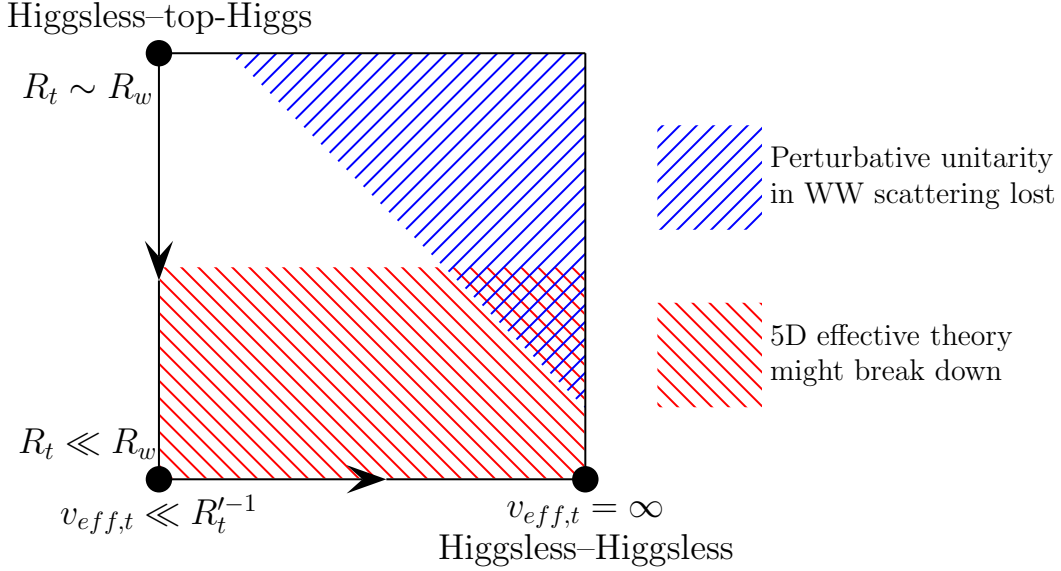}
\end{center}
\caption{\small A visualization of constraints on the parameter space. The dot at the upper-left is the 
higgsless---top-Higgs theory in which only the top Yukawa is large. 
The dot at lower right is the higgsless---higgsless theory we would like to ideally reach to decouple all scalars from the SM 
fields. Moving along the arrow pointing right, from higgsless---top-Higgs to higgsless---higgsless, one can potentially run into 
perturbative unitarity breakdown. This is not a danger when $R_t \ll R_w$, but as one moves along the downward arrow toward small 
$R_t/R_w$, one faces increasingly strong coupling among all KK modes on the new side. This signals 
a potential breakdown of the 5D effective theory.} \label{fig:schematic}
\end{figure}

Alternatively, we can consider the higgsless---top-Higgs model, in which we can 
take $f^2_{\pi, w} \approx f^2_{\pi, t}$ but $v_t$ small. Then we again find that the new side contributes little to electroweak 
symmetry breaking and perturbative unitarity is safe, and we also have a reasonable 5D effective theory. In this case the only 
strong coupling is a large Yukawa coupling for the top quark. Of course, such a model is not genuinely ``higgsless" 
in the sense that for small $v_t$ the Higgs on the new side does not decouple from the SM fields. 
However, this surviving Higgs will 
have small couplings to the Standard Model gauge bosons and large coupling to the top, so it can be unusually heavy and will have 
interesting and distinct properties. 
We show a summary of the different constraints on the parameters $R_t / R_w$ and $v_t$ (assuming $1/R_t^\prime$ 
large) in Fig. \ref{fig:schematic}.

\subsection{Phenomenology of the higgsless---top-Higgs model}

  From the previous discussions we can see that the model that is mostly under perturbative control is the higgsless---top-Higgs
model with $R_t \sim R_w$, $v_{{\rm \it eff},t} \lesssim 50$ GeV and $1/R_t^\prime$ at a scale of 2 to 5 TeV. In the following we will be discussing
some features of this model in detail.
\begin{figure}[tb]
\begin{center}
\includegraphics[width=6in]{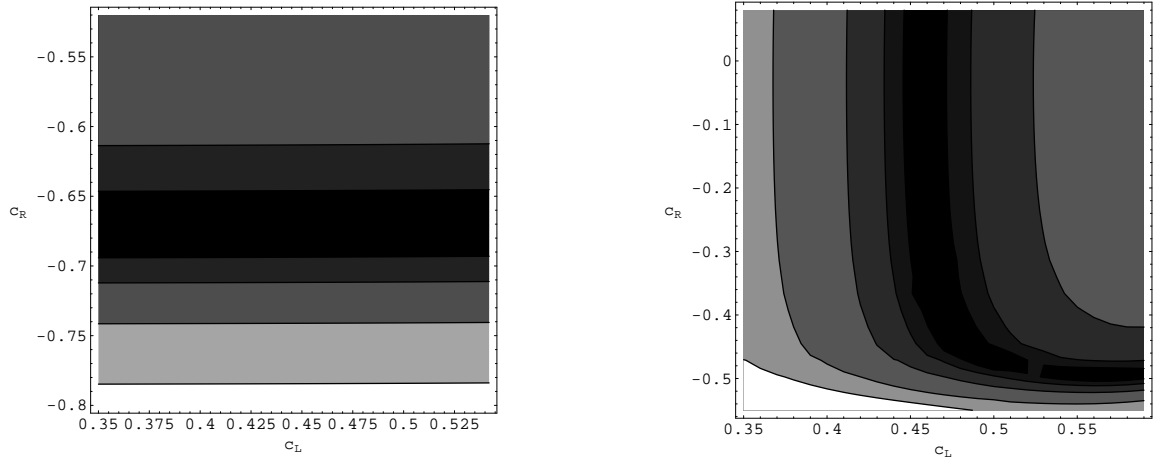}
\caption{\small Deviation of $Z b_l \bar{b}_l$ from SM value, as a function of bulk mass parameters, in the 
higgsless---top-Higgs case in the plot on the left and in the higgsless---higgsless case on the right. The coupling decreases from bottom-to-top in the left plot and left-to-right in the right plot. The 
contours (darkest to lightest) are at .5\%, 1\%, 2\%, 4\%, 
and 6\%.}\label{fig:Zbb}
\end{center}
\end{figure}

We can place the third-generation fermions 
in the new bulk and give them masses by Yukawa coupling to the brane-localized top-Higgs. The wave functions of the $W$ and $Z$ in the 
new bulk will be approximately flat. Then, from the perspective of third-generation physics (quantities like $m_t$ and the 
$Zb\bar{b}$ coupling), the physics in the new bulk looks essentially the same as that of the usual Randall-Sundrum model with 
custodial symmetry \cite{ADMS}, with two important differences. The first difference is that the top-Higgs VEV $v_t$ is small, so that the top 
Yukawa coupling must be large. The second is the presence of the top-pion scalar modes noted in the last section. Aside from this, 
the results must be much as in the usual Randall-Sundrum model. We find that we do not have to take either the left- or right-handed top quark extremely close to the TeV brane to obtain the proper couplings. The right-handed bottom quark 
will mix with Planck-brane localized fermions (or, alternatively, will have a large Planck-brane kinetic term) to split it from the 
top quark. The right-handed bottom then lives mostly on the Planck brane, and so will have the usual SM couplings. The problem 
arising in the original higgsless model was that a large mass $M_D R^\prime$ on the IR brane caused much of the left-handed 
bottom quark to live in the $SU(2)_R$ multiplet. Note that a similar problem would arise in a model with a brane-localized Higgs 
and the same value of $R^\prime$; the usual Randall-Sundrum models evade this problem with a large $1/R^\prime$. In our new 
scenario, $R_t^\prime$ is significantly smaller than $R_w^\prime$, so at tree level we are able to obtain the desired values of $m_t$ and the 
SM couplings of the bottom. We show this explicitly in Fig. \ref{fig:Zbb}. It corresponds to $R_w^{\prime -1}$ = 292 GeV, $R_t^{\prime -1}$ = 3 TeV, $R_w^{-1} = R_t^{-1} \approx 10^6$ GeV, $v_{{\rm \it eff},t}$ = 50 GeV, and a light reference fermion on the old side for which $c_L = .515$ (for these parameters we find $S \approx -.066$, $T \approx -.032$, and $U \approx .010$). Note that we can accommodate a change in the tree-level value in either direction, so loop corrections from the top-pion are not a grave danger. Furthermore, the masses of the lightest top and bottom KK modes are at approximately 6 TeV, so they should not cause dangerous contributions to the $T$ parameter.
\begin{figure}[tb]
\begin{center}
\includegraphics[width=7cm]{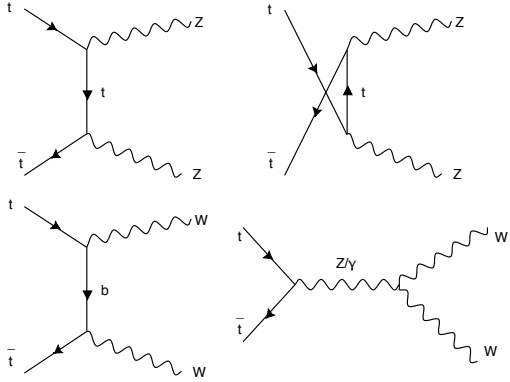}
\end{center}
\caption{\small Scattering processes for $t\bar{t} \rightarrow V_L V_L$ of top anti-top pairs into longitudinal vector bosons. These processes determine the unitarity bound on the mass of the heavy top-Higgs boson in the higgsless---top-Higgs model.} \label{fig:ttVV}
\end{figure}

We would like now to investigate the phenomenology of this model in more detail. 
A novelty with respect to the usual higgsless models is the presence of light scalars.
The top-Higgs on the AdS$_t$ side will couple strongly to tops and give a small contribution to the $W$mass, hence the name top-Higgs.
Its tree level mass is determined by the quartic coupling on the IR brane $m_h^2 = \lambda_t v_{{\rm \it eff},t}^2$, although large corrections could arise due to the strong coupling with the top.
An important parameter controlling this model is $\gamma = \sfrac{v_{{\rm \it eff},t}}{v}$, the ratio of the (warped down) top-Higgs VEV to the usual SM Higgs VEV. 
If $\gamma \ll 1$, we can have confidence that perturbative unitarity in $WW$ scattering is restored by the KK modes in the higgsless bulk, provided they have masses in the 700 GeV range. 
In this case, the top-Higgs is not needed for perturbative unitarity in $WW$ scattering, but there is still a unitarity bound on its mass. 
This bound arises from considering $t\bar{t} \rightarrow V_L V_L$ scattering, where $V_L$ denotes a longitudinal $W$ or $Z$ boson. 
The relevant tree-level diagrams are shown in Fig. \ref{fig:ttVV}. This bound sets the scale of new physics, $\Lambda_{NP}$, to be (as in \cite{TopHiggsUnitarity})
\beq
\Lambda_{NP} \leq \frac{4 \pi \sqrt{2}}{3 G_F m_t} \approx 2.8\ {\rm TeV}.
\eeq
This is computed in an effective theory given by the Standard Model with the Higgs boson removed and the Yukawa coupling of the top replaced with a Dirac mass. 
The resonances that couple to the top in our model are predominantly those on the new side, which have a mass set by $R_t^{\prime -1}$, which is large. 
Thus these resonances make little contribution to the scattering in question, and we can view $\Lambda_{NP}$ as a rough upper bound on the possible mass of the top-Higgs boson in our model. 
It is clear anyway that a heavy top-Higgs is allowed.
The other set of scalars is an $SU(2)$ triplet of pseudo-Goldstone bosons that we called top-pions.
We estimated their mass in Eq.\ref{eq:mA5}.
It is set by the scale $1/R_t^\prime$, so that we expect them to be quite heavy too, at least heavy enough to avoid direct bounds on charged scalars.
In the following we will assume a wide range of possibilities for the scalar masses, although the agreement of the $Z b \bar b$ coupling would suggest that they are heavy, likely above the $t \bar t$ threshold.

\begin{table}[t]
\centering
\begin{tabular}{|c|c|c|l|}
\hline
Decay Mode & $M_{ht} \gg M_{\pi t}$ & $M_{ht} \ll M_{\pi t}$ & Remarks \\
\hline
$t\bar{t}$ & 34\% & 98\% & {\small Large background, but consider associated $t\bar{t}H$.}\\
$W^{\pm} \pi_t^{\mp}$ & 43\% & -- & {\small Similar to $t\bar{t}$.}\\
$Z \pi_t^0$ & 22\% & -- & {\small Interesting. $t\bar{t}Z$: four leptons, two $b$ jets.}\\
$W^+ W^-$ & .35\% & 1.0\% & {\small Rare and probably difficult.}\\
$ZZ$ & .17\% & .50\% & {\small Usual ``golden" mode, but very rare.}\\ 
$gg$ & .06\% & .16\% & \\
$b\bar{b}$ & .01\% & .03\% & \\
$\gamma \gamma$ & $2.1\times 10^{-4}$ \% & $6.1 \times 10^{-4}$ \% & {\small Very rare, but sometimes accessible.} \\
\hline
\end{tabular}
\caption{\small Leading branching ratio estimates (subject to possibly order 1 corrections) for the heavy top-Higgs (assuming $M_{ht} \approx$ 1 TeV and $\gamma \approx 0.2$). In the case $M_{ht} \gg M_{\pi t}$, we have assumed $M_{\pi t} =$ 400 GeV for the purpose of calculation. These can receive large corrections, but the qualitative hierarchy (associated with $\gamma = \sfrac{v_{{\rm \it eff},t}}{v}$) should persist. Using the Pythia cross-section $\sigma \approx 88$ fb for a 1 TeV Higgs, rescaled by a factor of $\gamma^{-2} = 25$ to take into account enhanced production, we find an estimate of $\approx 1000$ $ZZ$ events in 100 fb$^{-1}$, but only about 1 $\gamma \gamma$ event. However, for $M_{ht} \approx 500$ GeV, we expect a larger cross section, $\approx \gamma^{-2} \times 1700$ fb, and there could be about 100 $\gamma\gamma$ events in 100 fb$^{-1}$. Note that the branching ratio estimates for the neutral top-pion will be essentially the same (with $M_{\pi t}$ and $M_{ht}$ reversed in the above table).}
\label{tab:HiggsBR}
\end{table}

\begin{table}[t]
\centering
\begin{tabular}{|c|c|c|l|}
\hline
Decay Mode & BR & Events in 100 fb$^{-1}$ & Remarks \\
\hline
$W^+ W^-$ & 40\% & $4.0 \times 10^6$ & Probably difficult.\\
$b\bar{b}$ & 22\% & $2.2 \times 10^6$ & Large QCD background. \\
$gg$ & 20\% & $2.0 \times 10^6$ & Large QCD background. \\
$ZZ$ & 18\% & $1.8 \times 10^6$ & Usual ``golden" mode, now rarer.\\ 
$\gamma \gamma$ & .07\% & 7000 & Light Higgs ``golden" mode, still visible.\\
\hline
\end{tabular}
\caption{\small Leading branching ratio estimates (subject to possibly order 1 corrections) for the top-Higgs when $M_{ht}$ is below the $t\bar{t}$ threshold and also below the top-pion threshold. These are calculating from rescaling the SM branching ratios using $M_{ht} \approx$ 300 GeV. The number of events is estimated via the Pythia cross-section, $\sigma = 3.9$ pb for $M_{ht}$ = 300 GeV, rescaled by a factor of $\gamma^{-2} = 25$ to take into account enhanced production. Alternatively, these can be viewed as approximate branching ratios of the neutral top-pion when its mass is below the $t\bar{t}$ threshold.}
\label{tab:pionBR}
\end{table}

The phenomenology of this model is still characterized by the presence of $W$and $Z$ resonances that unitarize the $WW$ scattering amplitude: this sector of the theory is under perturbative control so it is possible to make precise statements.
For a detailed analysis of the collider signatures of higgsless models, see ref.~\cite{BMP}.
Although the strong coupling regime does not allow us to make precise calculations, the presence of top-Higgses can provide interesting collider phenomenology for these models.
The case of a top-Higgs has already been considered in the literature in more traditional scenarios~\cite{Hill:topcolor,topHiggs,topHiggsII}.
The key feature of this model is the strong coupling with the top, determined by the large Yukawa coupling $~ m_t/v_{{\rm \it eff},t}$ that is enhanced by a factor $1/\gamma=v/v_{{\rm \it eff},t}$ with respect to the SM Higgs case.
On the other hand, the couplings with massive gauge bosons are suppressed by a factor $\gamma$, as the contribution of the top-Higgs sector to electroweak symmetry breaking is small.
There will be a coupling with the bottom, suppressed by $m_b/m_t$ with respect to the top coupling.
We also have couplings $h_t W^{\pm \mu} \pi^{\mp}_{t}$ and $h_t Z\pi^0_t$ of the top-higgs with SM gauge bosons and the top-pions. These arise from a term $2 g h_t A^a_{\mu} \partial^{\mu} \pi^a_t$ in the Lagrangian.

Let us first discuss the decays of the neutral scalars $h_t$ and $\pi_t^0$.
If their mass is above the $t\bar t$ threshold, they will often decay into tops. Notably, if the mass is large, say 1 TeV, multiple top decays will not be suppressed due to the strong coupling. However, for $M_{ht} \gg M_{\pi t}$, the width of the cascade decay $h_t \rightarrow W^{-} \pi^{+}_t$ is $\sfrac{g^2}{16 \pi} \sfrac{M^3_h}{M^2_W}$, becoming quite large for a very heavy top-Higgs, and even surpassing the enhanced decay to tops. There is a suppressed tree-level decay to the weak gauge bosons.
Virtual tops will also induce loop decays into gauge bosons, $\gamma \gamma$, $W^+ W^-$, $Z Z$, $g g$, and we generically expect them to be suppressed by a factor of about $\left(\sfrac{\alpha}{3\pi}\right)^2$ or $\left(\sfrac{\alpha_s}{3\pi}\right)^2$ with respect to the $t \bar t$ channel. A simple estimate shows that the $\gamma \gamma$ decay is suppressed, but it could still be present in a measurable number of events at the LHC in decays of the neutral top-pion or even of the top-Higgs if it is not too heavy. We summarize the various modes in Tables \ref{tab:HiggsBR} and \ref{tab:pionBR}. The widths are calculated at leading order. That is, the tree-level decays are calculated by rescaling the tree-level SM widths by appropriate powers of $\gamma$, except for the decays through a top-pion, which are computed directly in our model. (The decay to $b\bar{b}$ is computed with the running $b$ quark mass.) Loop-level decays are calculated by rescaling one-loop SM results by appropriate powers of $\gamma$. These estimates should provide the right qualitative picture, though the large couplings of the top could induce order one changes.
Other interesting channels could open up if we consider flavor violating decays, like for example 
$t \bar{c}$. It might be relevant or even dominant in the case of a relatively light scalar (below the $t \bar{t}$ threshold)~\cite{topHiggsII}. However, these channels are highly model dependent: in this scenario the flavor physics is generated on the Planck brane via mixings in non-diagonal localized kinetic terms. For instance, we can choose the parameters so that there is no $\pi_t t \bar{c}$  coupling at all. So, we will not  consider this possibility further.   Regarding the charged top-pion, it will mostly decay into $t b$ pairs, though at loop level there will be rare decays to $W^{\pm} \gamma$ and $W^{\pm} Z$, which could lead to interesting signatures.

\begin{figure}[tb]
\begin{center}
\vspace{.5cm}
\includegraphics[width=12cm]{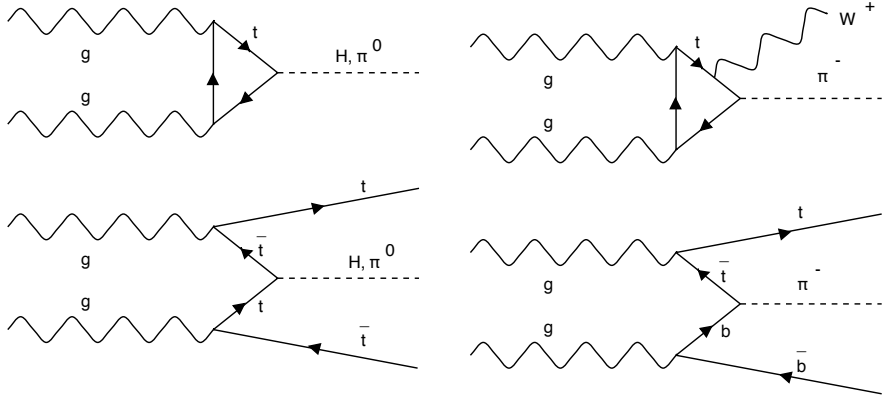}
\end{center}
\caption{\small Gluon-gluon fusion processes producing top-higgs and top-pion bosons at the LHC.} 
\label{fig:higgsprod}
\end{figure}

At the Large Hadron Collider (LHC) we expect a lot of top-Higgses and top-pions to be produced (see Fig.~\ref{fig:higgsprod}), via the usual gluon fusion or top fusion, now enhanced with respect to the SM one by the large Yukawa coupling.
If the mass is larger than $2 m_t$, the main decay channel is in $t \bar t$, or multiple tops.
The QCD background is large, however it is probably realistic to search the spectrum of the $t\bar t$ events due to the enhanced production rate in gluon-fusion.
A golden channel is represented by the decay into two photons or two $Z$ bosons. We expect a substantial number of $ZZ$ events throughout a wide range of masses. To observe $\gamma \gamma$ events, which are enhanced relative to the SM by the large Yukawa and by the enhanced production of neutral top-pion and of top-Higgs, we need relatively small masses. At $M_{ht} \approx$ 1 TeV, the cross-section is expected to be too small to observe a substantial number of events. We have used Pythia \cite{pythia} to estimate the SM cross-section for a Higgs produced by gluon-gluon fusion. This cross-section, suitably enhanced by the large Yukawa, was used to estimate numbers of LHC events per 100 fb$^{-1}$ in Tables \ref{tab:HiggsBR} and \ref{tab:pionBR}. Of course, strong coupling will modify our estimates of cross-sections and branching ratios, so the numbers we present should be taken as order-of-magnitude guides. We expect the neutral top-pion to have a mass somewhat below the TeV scale, so optimistically one should see the photon-photon channel from the neutral top-pion irrespective of the top-Higgs mass.  We stress that, for a mass in the 500 GeV region, one can expect roughly one photon-photon event per fb$^{-1}$, while the number of $ZZ$ events should be of order one thousand times larger.
At the LHC it will be relatively easy to see peaks in the two photons or $\ell^+ \ell^- \ell^+ \ell^-$ channels, due to the reduced background.
Thus, a heavy resonance in $\gamma \gamma$, associated with an anomalous production of multi-top events would be a striking signature of these models.
If cascade decays of the top-Higgs into the top-pion are allowed, we could also observe interesting $Z t \bar t (t\bar t)$ channels that could lead to striking 6 leptons 4 $b$ events.
If the masses are below the top threshold, the main decay channels will be into $b$'s and gauge bosons.
The golden channels are again $\gamma \gamma$ and $Z Z$. The high rate of $b\bar{b}$ events even above the $WW$ threshold could help distinguish a light scalar in our model from a heavy SM Higgs. Also, if the rate of $\gamma \gamma$ is not too far below the rate of $\ell^+ \ell^- \ell^+ \ell^-$ events it would suggest a large top-loop induced coupling, since in the SM this ratio is fixed to be roughly $\left(\sfrac{\alpha}{3\pi}\right)^2$.

Finally, the charged top-pion would be harder to study: its production is suppressed as we do not have a gluon fusion channel producing solely a top pion. It will be produced in association with a $t b$ pair or with a $W$ boson.
It will then most likely decay into $t b$, so that its signal will suffer from a large QCD pollution.
An interesting effect could be an anomalous production of multi $b$-jet events. The loop-level decays to $W\gamma$ and $WZ$ could produce interesting multi-gauge-boson events, but these have a suppression comparable to the $\gamma\gamma$ decays of the neutral top-pion.

\subsection{Phenomenology of the higgsless---higgsless model}

Finally, we summarize the tree-level numerical results for the higgsless---higgsless limit. Let us first discuss how to
fix the values of the parameters corresponding to a potentially interesting theory. First of all, we would like 
one of the sides to be a higgsless model as in~\cite{CuringIlls}, with low enough KK masses for the gauge bosons to 
ensure perturbative unitarity of the $WW$scattering amplitudes. This can be achieved if the first resonance mass is around 600-700 GeV, thus fixing $R_w^{\prime -1} \sim 300$ GeV.
The value of the $W$ mass will fix the $\log R_w'/R_w \sim 10$, so that a natural value for $R_w^{-1}$ is around $10^8$~GeV.
On the new side we need the IR scale to be large enough to accomodate the top mass, so that $R_t^{\prime -1}\sim 2-5\ TeV \gg R_w^{\prime -1}$.
However, we want to do it without a low-scale violation of perturbative unitarity.
Since the KK modes on the new side will be very heavy 
$>$ TeV, this is only possible if the new side does not contribute a lot to the $W$ mass itself. 
From (\ref{Wmasssimple}) we can see that this can be achieved by choosing a smaller curvature radius for the new side  $R_t \ll R_w (R'_t/R'_w)^2$.
For simplicity 
we will also assume the 5D gauge couplings are the same in both bulks, and that $g_{5L} = g_{5R}$.
Then, for any given value of $R'_t$ and of the contribution of the new side to the $W$ mass (that will determine the perturbative unitarity breakdown scale), we determine the scales $R_{w,t}$ and $R'_w$, while $\tilde{g}_5$ is fixed by the $Z$ mass.

We also choose a ``reference" bulk fermion in the old bulk as in \cite{CuringIlls} to fix the wave-function normalizations. 
As in the one-bulk case, when this reference fermion has an approximately flat wave-function ($c_L \sim 0.5$) the tree-level 
precision electroweak parameters $S$, $T$, and $U$ can all be made small.

To fix the actual numerical values we choose $R_t^{\prime -1} = 3$ TeV, and we allow the new bulk to contribute 5\% of the $W$ mass,
 $f^2_{\pi, t} \approx 0.05 (f^2_{\pi, w}+f^2_{\pi, t})$.
Then for the values quoted above, we get $R_w^{\prime -1} \approx 276$~GeV, and the curvature scales are $R_w^{-1} \approx \times 10^{8}$~GeV and $R_t^{-1} \approx 2 \times 10^{11}$~GeV. 
Choosing our reference quark to be a massless left-handed quark with $c_L = 0.46$, we find at tree-level 
$S \approx -0.08$, $T \approx -0.04$, and $U \approx 0.01$. 

The first task after fixing the parameters is to verify that the scale of perturbative unitarity violation is indeed 
pushed above the usual SM scale of 1 TeV. For this we can study 
the sum rules \cite{CGMPT}
 that the KK modes masses and couplings have to satisfy in order for terms in the scattering amplitudes that 
grow with a powers of the energy to cancel. We solve numerically for the Kaluza-Klein 
resonances of the $Z$ boson. The first one is at $M_Z^{(1)} \approx 676$~GeV $\approx 2.45 R_w^{\prime -1}$, as expected. Summing the KK 
modes up to 8 TeV, we find that the $E^4$ sum rule is satisfied to a precision of $2\times10^{-6}$ and the $E^2$ sum rule to a 
precision of $5\times10^{-3}$. 
\begin{figure}[tb]
\begin{center}
\includegraphics[width=11cm]{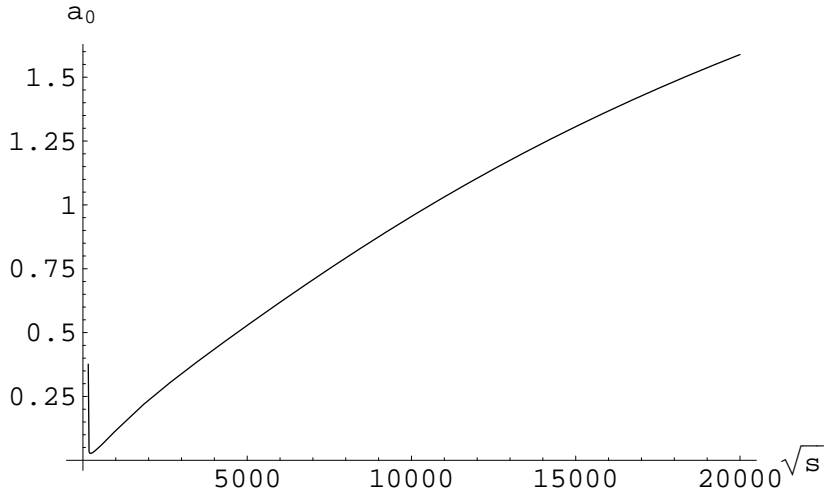}
\end{center}
\caption{\small Examining perturbative unitarity: the leading partial-wave amplitude $a_0$, as a function of 
center-of-mass energy.} \label{fig:unitarity}
\end{figure}
In order to find the unitarity violation scale we have shown in 
Fig. \ref{fig:unitarity} the s-wave partial-wave amplitude $a_0$ as a function of energy, which is obtained by 
numerically solving for all KK mode masses and couplings below 8~TeV, and then approximating the rest of the tower by an additional heavy mode so the graph does not misbehave at high energy (note that this approximation has no effect below 8 TeV). We can see that the unitarity bound from $a_0$ is around 5~TeV, well 
above the SM scale, and a scale likely inaccessible to the LHC. As explained in \cite{Papucci}, we should only rely on the low-energy linear behavior of this function,
which tells us that the effective theory is valid up to 5~TeV.

After fixing the parameters in the gauge sector, we are finally ready to consider the physics of the third generation
quarks. These particles are assumed to live on the new side, but the mass generation mechanism for them 
would be just like for the other fermions: a Dirac mass $M_D$ on the new IR brane would give a common mass to 
top and bottom, and the bottom mass would then be suppressed by a large kinetic term on the Planck brane for $b_R$ 
whose coefficient is $\xi_b$. We can then proceed in the following way:
for a given choice of bulk masses $c_L$, $c_R$, we can solve for the requisite Dirac mass $M_D$ to get the correct top mass $m_t$, 
and then for the mixing $\xi_b$ needed on the Planck brane to get the correct bottom mass. 

We can then numerically find the $Zb\bar{b}$ coupling as a function of $c_L,c_R$. 
We show a plot of the deviation of the $Z b_l \bar{b}_l$ coupling from the Standard Model value in 
Fig.~\ref{fig:Zbb}. Note that 
there is a band, where $c_L \approx 0.46$, where for a wide range of choices of $c_R$, $Z b_l \bar{b}_l$ is consistent with the SM 
value. This exactly corresponds to picking $c_L$ equal to the reference value of the light fermions on the old side.
On one side of the band the coupling  is larger, and on the other side it is smaller. Thus a wide range of loop corrections
to the $Zb_l\bar{b}_l$ coupling from the top-pion contribution can be accommodated in this model by changing the values of 
$c_{L,R}$, and tuning the sum of the tree-level plus loop corrections to equal the SM value. 
Thus we conclude that while in this model there is no {\it a priori} reason to expect this coupling to take on its SM value, 
parameters can likely be chosen such that the SM value could be accommodated. 

Since we cannot calculate the loop corrections, for concreteness
we will examine a case in which the tree-level value of the $Z b_l \bar{b}_l$ coupling agrees with the SM. We take 
$c_L = 0.46$, $c_R = -0.1$. We then find that we need to take $M_D \approx 610$~GeV and $\xi_b \approx 6000$ to obtain 
$m_t = 175$~GeV and $m_b = 4.5$~GeV. The tree-level $Z b_l \bar{b}_l$ coupling then deviates from the SM value by only 
$.03\%$. We calculate now the various couplings of the pseudo-Goldstones to the top and bottom. We find that the couplings 
involving the right-handed bottom are small: $g_{\pi_t^0 \bar{b}_l b_r} \approx g_{\pi_t^+ \bar{t}_l b_r} \approx -0.106$. 
However, as expected, the couplings involving the right-handed top are large: $g_{\pi_t^0 \bar{t}_l t_r} 
\approx g_{\pi_t^- \bar{b}_l t_r} \approx -4.16$.  Thus the top-pion coupling is four times larger than the SM Higgs coupling.

\section{Conclusions}
\setcounter{equation}{0}
\setcounter{footnote}{0}

We have considered extra dimensional descriptions of topcolor-type models. From the 4D point of view
these would correspond to theories where two separate strongly interacting sectors would contribute to 
electroweak symmetry breaking. In the 5D picture these would be two separate AdS bulks with their own IR 
branes, and the two bulks intersecting on the common Planck brane. The motivation for considering such models
is the need to separate the dynamics that gives most of electroweak symmetry breaking from that responsible 
for the top quark mass (which is the main problematic aspect of higgsless models of electroweak symmetry breaking).

We have described how to find the appropriate matching and boundary conditions for the fields that propagate 
in both sides, and gave a description of electroweak symmetry breaking if both IR branes have localized Higgs fields.
We have considered both the cases when the Higgs VEVs are small or large (the higgsless limit). We discussed the 
CFT interpretation of all of these limits, and also showed that a light pseudo-Goldstone boson (``top-pion'')
has to emerge in these setups. Depending on the limit considered, the top-pion could be mostly contained in one of 
the brane Higgses or in $A_5$ in the higgsless limit. 

Finally, we have used these models to try to resolve the issues surrounding the third generation quarks in the higgsless 
theories. In these models one of the bulks is like a generic higgsless model as in~\cite{CuringIlls} with only the 
light fermions propagating there, while the new bulk will contain the top and bottom quark, but will not be the 
dominant source of electroweak symmetry breaking. The suppression of the contribution to the $W$ mass  from the new 
side is either obtained by a small top-Higgs VEV (higgsless---top-Higgs models) or via a small curvature radius in the new bulk. 
A generic issue in all of these cases will be that the top-pions (and eventually the top-Higgs) are strongly coupled to the top and bottom quarks.
In the higgsless---higgsless case the small curvature radius will also imply that the KK modes dominantly living on the 
new side will be strongly coupled among themselves. In both limits the tree-level top mass and $Zb\bar{b}$ couplings can be made
to agree with the experimental results, however, due to the coupling of the top-pion one also needs to worry about 
large shifts from loop corrections. We have discussed the basic phenomenological consequences of both limits.
The top-pion and top-Higgs are expected to be largely produced at LHC.
Their main signature would be an observable heavy resonance in the $\gamma \gamma$ channel in association with an anomalously large rate of multi-top events.

\section*{Acknowledgments}
We thank Kaustubh Agashe, Gustavo Burdman, Roberto Contino, Josh Erlich, Guido Marandella and Alex Pomarol for useful discussions and comments.
J.T. thanks the KITP for its hospitality while part of this work was completed.
C.C. thanks the SPhT at CEA/Saclay for its hospitality while this work was in preparation.
The research of G.C. and C.C.
is supported in part by the DOE OJI grant DE-FG02-01ER41206 and in part
by the NSF grants PHY-0139738  and PHY-0098631. M.R. is supported by a Cornell University Graduate Fellowship.
C.G. is supported in part by the RTN European Program
MRTN-CT-2004-503369 and by the ACI Jeunes Chercheurs 2068. J.T. is supported
by the US Department of Energy under contract
W-7405-ENG-36 and in part by the National Science Foundation under grant No. PHY99-07949.

\end{document}